\documentclass[12pt]{article}

\usepackage{scicite}
\usepackage{times}
\usepackage{graphicx}
\usepackage[utf8]{inputenc}
\usepackage[T1]{fontenc}
\usepackage[table]{xcolor}
\usepackage{booktabs}
\usepackage{url}
\usepackage{comment}

\usepackage{amsmath}
\numberwithin{equation}{section}

\usepackage{caption}
\newenvironment{sciabstract}{%
\begin{quote} \bf}
{\end{quote}}

\title{A compositional link between rocky exoplanets \\ and their host stars} 

\author
{\normalsize Vardan Adibekyan,$^{1,2\ast}$ Caroline Dorn,$^{3}$ S\'{e}rgio G. Sousa,$^{2}$
Nuno C. Santos,$^{1,2}$  Bertram Bitsch,$^{4}$ \\ \normalsize Garik  Israelian,$^{5,6}$ Christoph Mordasini,$^{7}$
Susana C. C. Barros,$^{1,2}$   Elisa Delgado Mena,$^{1}$ \\ \normalsize Olivier D. S. Demangeon,$^{1,2}$
Jo\~{a}o P. Faria,$^{1}$ Pedro Figueira,$^{8,1}$  Artur A. Hakobyan,$^{9}$ \\ \normalsize
Mahmoudreza Oshagh,$^{5,6}$ B\'{a}rbara M. T. B. Soares,$^{1,2}$   Masanobu Kunitomo,$^{10}$ \\ \normalsize  Yoichi Takeda,$^{11,12}$
Emiliano Jofr\'{e},$^{13,14,15}$ Romina Petrucci,$^{14,15}$ Eder Martioli,$^{16,17}$\\
\\
\footnotesize{$^{1}$Instituto de Astrof\'isica e Ci\^encias do Espa\c{c}o, Universidade do Porto, }\\
\footnotesize{Centro de Astrof\'{\i}sica da Universidade do Porto, 4150-762 Porto, Portugal}\\
\footnotesize{$^{2}$Departamento de F\'{\i}sica e Astronomia, Faculdade de Ci\^encias, Universidade do Porto, 4169-007 Porto, Portugal}\\
\footnotesize{$^{3}$University of Zurich, Institut of Computational Sciences, CH-8057, Zurich, Switzerland}\\
\footnotesize{$^{4}$Max-Planck-Institut f\"{u}r Astronomie, 69117 Heidelberg, Germany}\\
\footnotesize{$^{5}$Instituto de Astrof\'{i}sica de Canarias, E-38205 La Laguna, Tenerife, Spain}\\
\footnotesize{$^{6}$Departamento de Astrof\`{i}sica, Universidad de La Laguna, E-38206 La Laguna, Tenerife, Spain}\\
\footnotesize{$^{7}$Physikalisches Institut, University of Bern, 3012 Bern, Switzerland}\\
\footnotesize{$^{8}$European Southern Observatory, Alonso de Córdova 3107,}
\footnotesize{Vitacura, Región Metropolitana, Chile}\\
\footnotesize{$^{9}$Center for Cosmology and Astrophysics, Alikhanian National Science Laboratory, 0036 Yerevan, Armenia}\\
\footnotesize{$^{10}$Department of Physics, School of Medicine, Kurume University, Fukuoka 830-0011, Japan}\\
\footnotesize{$^{11}$National Astronomical Observatory, Mitaka, Tokyo 181-8588, Japan}\\
\footnotesize{$^{12}$S\={o}g\={o} kenky\={u} daigakuin daigaku, The Graduate University for Advanced Studies, Mitaka, Tokyo 181-8588, Japan}\\
\footnotesize{$^{13}$Instituto de Astronom\'ia, Universidad Nacional Aut\'onoma de M\'exico,}\\
\footnotesize{Ciudad Universitaria, Ciudad de M\'{e}xico, C.P. 04510, M\'exico}\\
\footnotesize{$^{14}$Universidad Nacional de C\'ordoba - Observatorio Astron\'omico de C\'ordoba, X5000BGR C\'ordoba, Argentina}\\
\footnotesize{$^{15}$Consejo Nacional de Investigaciones Cient\'ificas y T\'ecnicas, Argentina}\\
\footnotesize{$^{16}$Institut d’Astrophysique de Paris, UMR7095 Centre national de la recherche scientifique,}\\
\footnotesize{Sorbonne Universit\'{e}, 75014 Paris, France}\\
\footnotesize{$^{17}$Laborat\'{o}rio Nacional de Astrof\'isica, Itajub\'{a} MG 37504-364, Brazil}\\
\footnotesize{$^\ast$E-mail:  vadibekyan@astro.up.pt}
}

\date{}

\begin{document} 


\baselineskip24pt


\maketitle 

\newpage


\begin{sciabstract}
  Stars and planets both form by accreting material from a surrounding disk. Because they grow from the same material, theory predicts that there should be a relationship between their compositions. We search for a compositional link between rocky exoplanets and their host stars. We estimate the iron-mass fraction of rocky exoplanets from their masses and radii and compare it with the compositions of their host stars, which we assume reflect the compositions of the proto-planetary disks. We find a correlation (but not a 1:1 relationship) between these two quantities, with a slope of $>$4, which we interpret as due to planet formation processes. Further, we find that super-Earths and super-Mercuries appear to be distinct populations in compositional space with implications to their formation.
\end{sciabstract}


\section*{}

Characterising the interiors of rocky exoplanets requires measurements of both their mass and radius. These are usually provided by a combination of two techniques: planetary radius is determined via transit observations and planetary mass via radial velocity (RV) measurements. The derivation of these parameters relies on properties of the host stars, so limited by the precision with which those are measured. Exoplanets orbiting stars which are similar to the Sun (F,G,K spectral types) benefit from the precise characterisation possible for their host stars\cite{Sousa-08}.

We consider a set of 32 low-mass exoplanets (planet mass $M$ $<$ 10 Earth masses, $M_{\mathrm{\oplus}}$) orbiting 27 F,G,K stars with uncertainties in both planet mass and radius of < 30\%\cite{supl}. Above $\sim$4 $M_{\mathrm{\oplus}}$, the distribution of these planets on a mass-radius diagram shows an apparent gap between two populations (Fig.~1). This gap has been proposed as the separation between rocky exoplanets and gas-rich mini-Neptunes \cite{Fulton-17,Owen-17,Ginzburg-18,Jin-18,Venturini-20}. We investigate how the properties of these rocky planets depend on the composition of their host stars. We therefore discard the planets above the radius gap in Figure~1, leaving 22 planets without substantial water or gas-rich envelopes i.e. rocky planets.

\begin{figure}[h!]
  \begin{minipage}[c]{0.6\textwidth}
    \includegraphics[width=\textwidth]{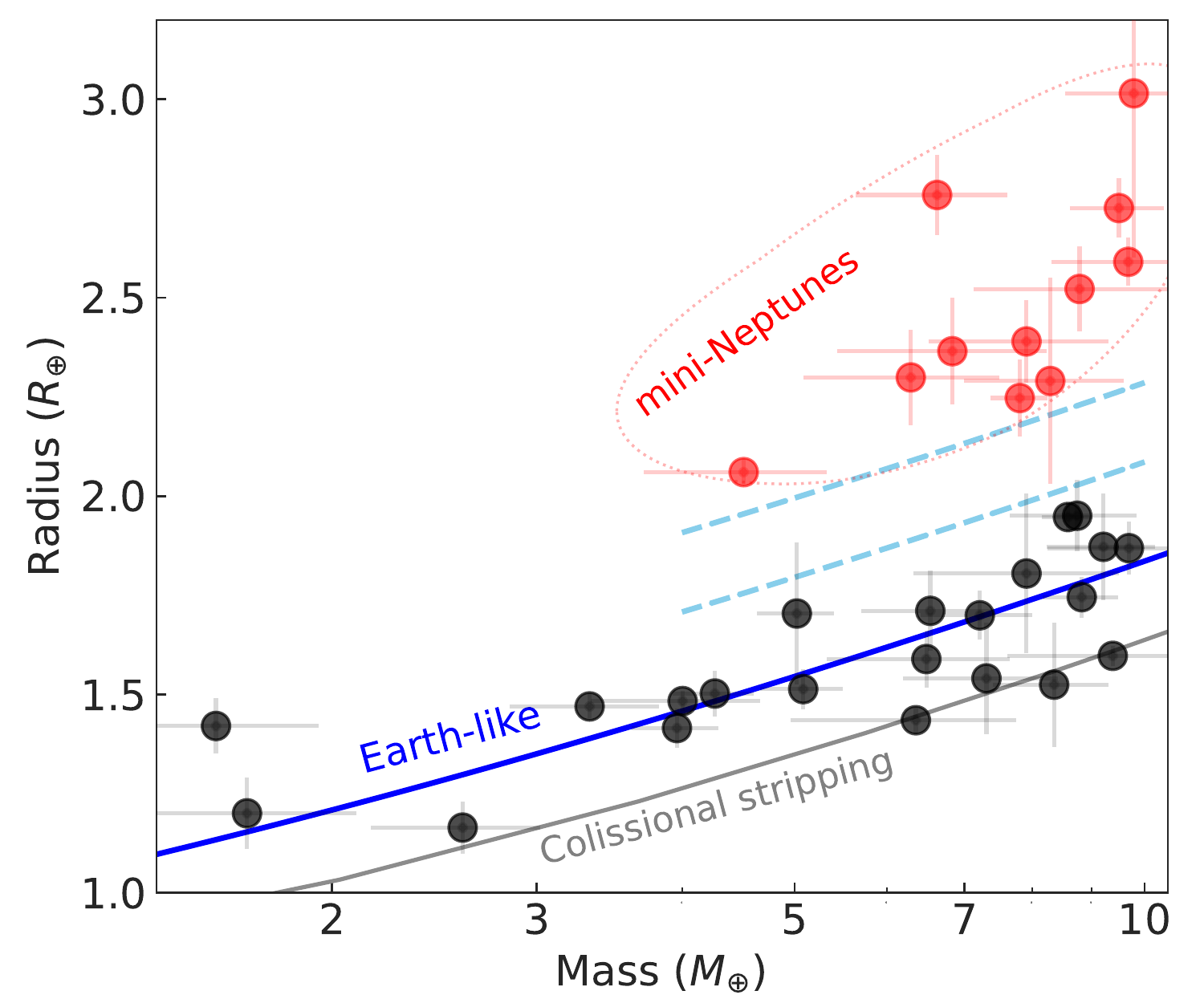}
  \end{minipage}\hfill
  \begin{minipage}[c]{0.37\textwidth}
    \caption{\footnotesize \textbf{Fig. 1. Mass-radius diagram for the rocky planets  in our sample.} All these planets have masses below 10 $M_{\mathrm{\oplus}}$ and uncertainty both in mass and radius $<$ 30\%. The radii of the planets are in Earth radius ($R_{\mathrm{\oplus}}$) The light-blue dashed curves, drawn by eye, indicate the location of the radius gap that separates the 'mini Neptunes' with gaseous envelopes (red circles) from the planets without gaseous envelopes (black circles). The solid blue curve shows the mass-radius relationship expected for Earth-like composition (32\% Fe + 68\% MgSiO3)\cite{Dorn17}. The grey solid curve marks the minimum planetary radius predicted by a collisional stripping  (giant impact) model\cite{Marcus-10}. All error bars show one standard deviation.} \label{fig:m_r_diagram}
  \end{minipage}
\end{figure}

The abundances of elements in the atmospheres of main-sequence stars reflects their bulk composition (except the lightest elements) within a few percent\cite{Dotter-17}, which meteorite measurements have shown to be valid for the Sun\cite{Asplund-09}. Theory predicts that the Fe/Si and Mg/Si abundance ratios in stars and planets remain very similar during the planet formation processes\cite{Bond-10,Thiabaud-14,Bonsor-21}. The atmospheric abundances of refractory elements (such as Mg, Si, and Fe) of solar-type stars are therefore considered a proxy of the composition of the initial protoplanetary disk\cite{Dorn-15,Unterborn-16}. 

We analysed spectra of the host stars of the 21 selected planets (HD\,80653 does not have an available spectrum) and measured their atmospheric chemical compositions\cite{supl}. We determined the abundances of Mg, Si, and Fe  in the host stars, which are the major rock-forming elements\cite{Lodders-03}. We then inserted these abundances into a simple stoichiometric model\cite{Santos-15,Santos-17} to estimate the iron-to-silicate mass fraction ($f_{\mathrm{iron}}^{\mathrm{star}}$) of proto-planetary disks. For the proto-solar disk, this model predicts a $f_{\mathrm{iron}}^{\mathrm{Sun}}$ of 33.2$\pm$1.7\%,  consistent with the iron content of the Earth ($\sim$32\%), Venus ($\sim$32\%), and Mars ($\sim$30\%), but different from the one of Mercury ($\sim$70\%)\cite{supl}.

\begin{figure}[h!]
  \begin{minipage}[c]{0.6\textwidth}
    \includegraphics[width=\textwidth]{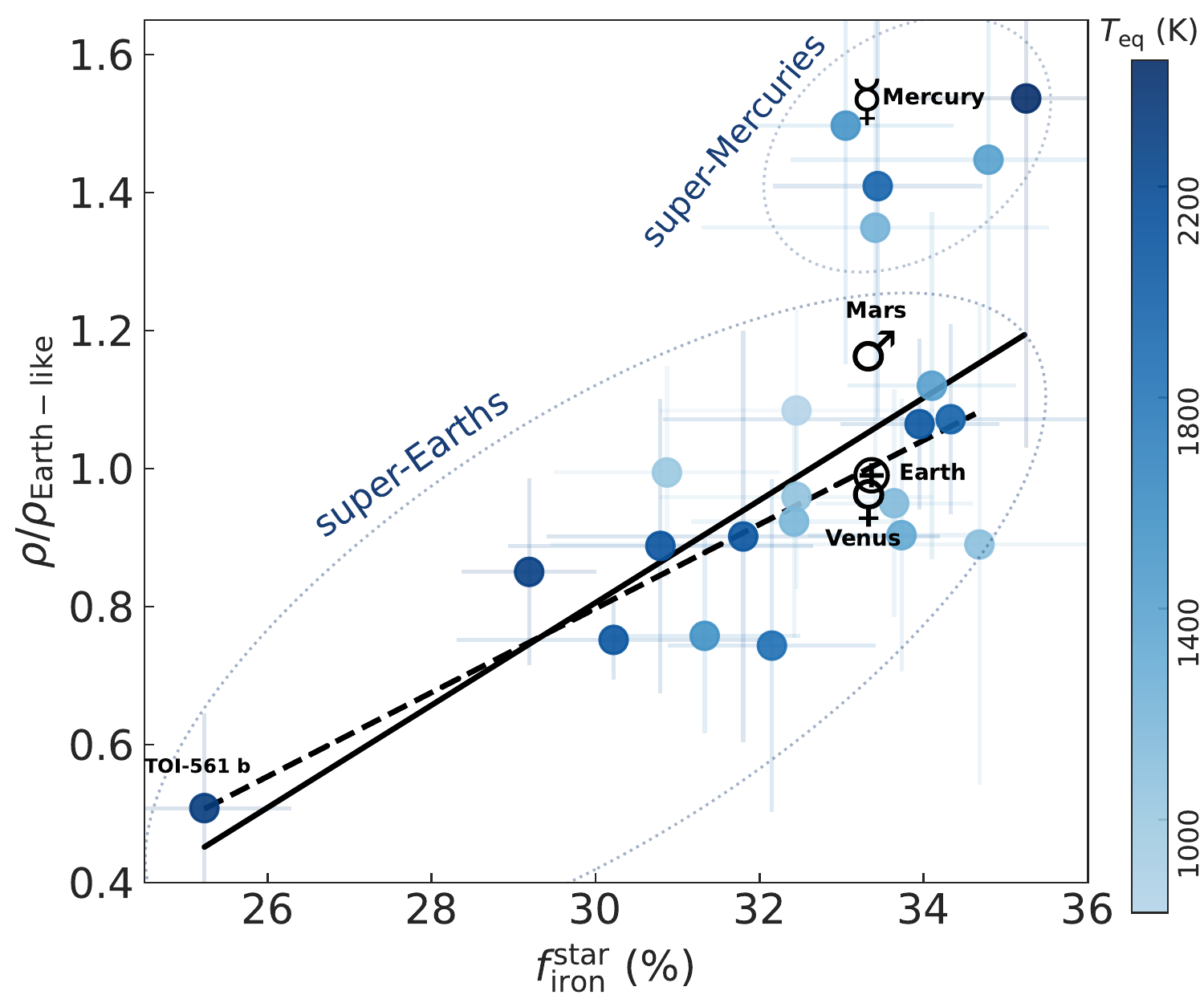}
  \end{minipage}\hfill
  \begin{minipage}[c]{0.38\textwidth}
    \caption{\footnotesize \textbf{Fig. 2. Densities of rocky planets as a function of iron fraction.} The measured density of each planet was normalised by the expected density of an Earth-like composition at that mass (\cite{Dorn17}, Fig. 1). The iron mass fraction was calculated from the elemental abundances of the host star. The rocky planets in the Solar System are indicated with their respective symbols in black; these all have the same iron fraction because this was derived from the abundances of the Sun, not measured directly from the planets. The black solid and dashed lines show the correlation for the full sample and for the sample after excluding the five potential super-Mercuries, respectively. The Solar System planets are not included in the linear regressions. The exoplanet symbols (blue circles) are color-coded by the equilibrium temperature of the planets to look for possible trends. No color gradient is observable. All error bars show one standard deviation.} \label{fig:density}
  \end{minipage}
\end{figure}

At a given mass, the rocky planets in our sample have a dispersion in radius around the  curve expected for Earth-like composition (Fig.~1). To investigate whether the densities of these planets are linked to the primordial composition of the protoplanetary disk, Fig.~\ref{fig:density} shows the planet density $\rho$, normalised to that expected for an Earth-like composition $\rho_{\mathrm{Earth-like}}$ \cite{Dorn17} as a function of $f_{\mathrm{iron}}^{\mathrm{star}}$. The normalization accounts for planets with different masses having different densities even with the same composition, due to compression. We find a relationship between these two quantities, indicating that the final planetary density is a function of the composition of the protoplanetary disk. We performed an orthogonal distance regression (ODR) to quantify the relation and then used $t$-statistics to assess its significance, finding that the observed correlation is statistically significant with a $p$-value of $\sim$ 7$\times$10$^{-6}$\cite{supl}.

Because the normalised density is calculated from the observed properties of the planets, while $f_{\mathrm{iron}}^{\mathrm{star}}$ is inferred from the host star composition, this trend provides  observational evidence for a compositional link between rocky exoplanets and their host stars. Rocky planets therefore preserve information about the composition of the proto-planetary disk within which they formed.

Although Fig.~\ref{fig:density} shows a correlation, it is evident that for a given $f_{\mathrm{iron}}^{\mathrm{star}}$ rocky planets can have a range of densities. The observed scatter is compatible with the average uncertainty of $\rho / \rho_{\mathrm{Earth-like}}$. We nevertheless tested whether part of this dispersion could have an astrophysical origin. The size and thus density of planets could be influenced by the flux of high-energy photons that planets receive from their host stars \cite{Jin-18}. Extreme irradiation from the host star can lead to atmospheric escape from low-mass planets with a substantial impact on the evolution of their bulk composition\cite{Owen-19}.  However, we found no correlation between the normalized density of the planets and their equilibrium temperature, $T_{\mathrm{eq}}$ (Fig.~\ref{fig:density}). Assuming that these planets have maintained a constant $T_{\mathrm{eq}}$ since their formation, this suggests that for rocky planets $T_{\mathrm{eq}}$ does not have a dominant impact on their radius. Alternatively, a radial oxidation gradient in the protoplanetary disks might lead to a correlation between orbital distance and composition, and therefore also between orbital distance and planet density\cite{Rubie-15}. However, we also found no correlation between $\rho / \rho_{\mathrm{Earth-like}}$ and the orbital distance of the planets (Fig. S1).

\begin{figure}[h!]
\begin{center}
\includegraphics[width=0.9\linewidth]{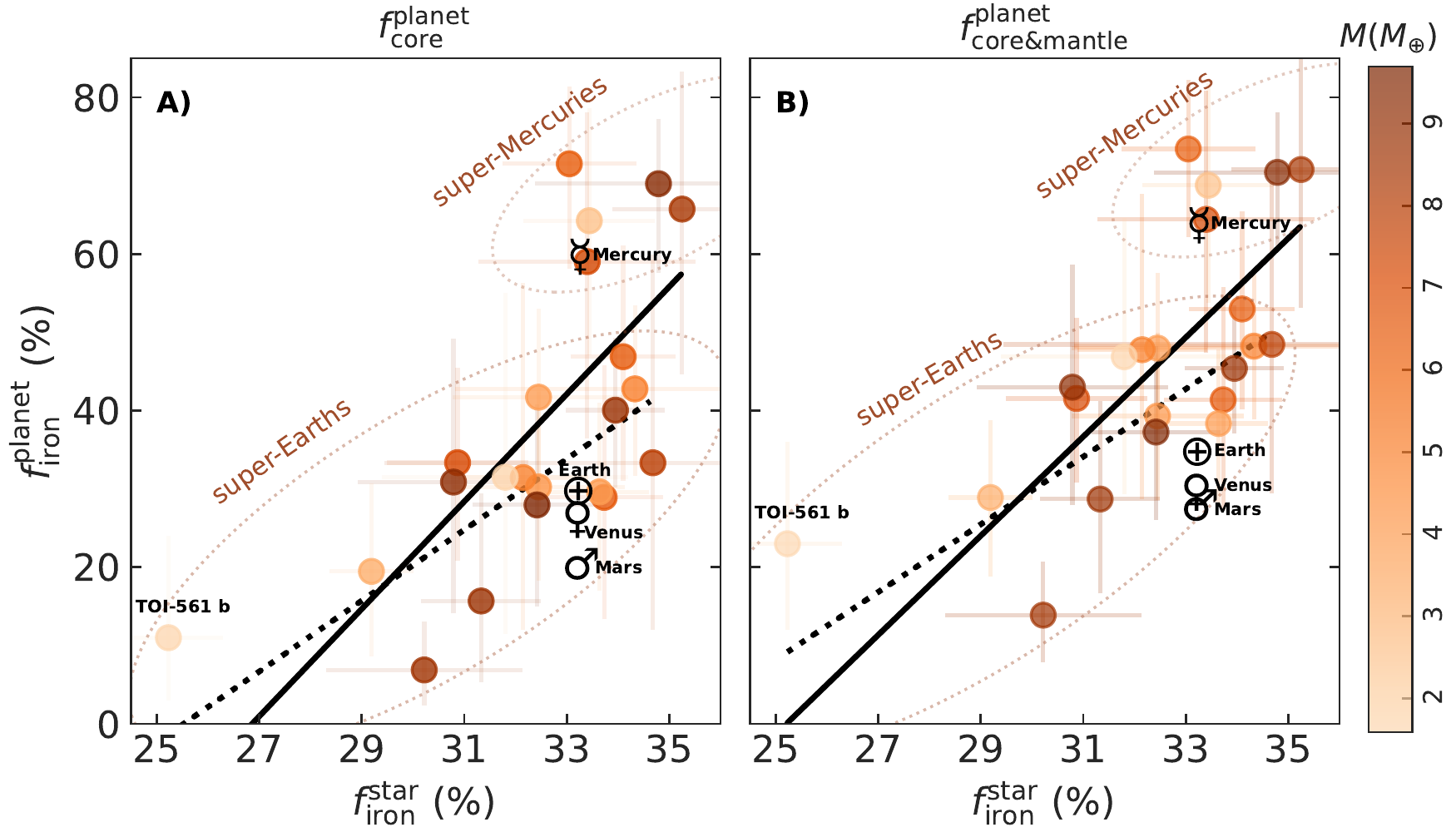}
\end{center}
\caption{\footnotesize \textbf{Fig. 3. Iron contents of rocky planets.} The iron mass fraction of the protoplanetary disk, estimated from the host star abundances ($f_{\mathrm{iron}}^{\mathrm{star}}$) is plotted as a function of iron mass fraction inferred from the planet's mass and radius ($f_{\mathrm{iron}}^{\mathrm{planet}}$) as in Figure 2. $f_{\mathrm{iron}}^{\mathrm{planet}}$ is shown for two assumptions: (A)  all iron resides in the core only or (B) iron present in both mantle and core. The color bar indicates the mass of the planets. Symbols and line styles are the same as in Figure 2. The error bars of $f_{\mathrm{iron}}^{\mathrm{star}}$ show one standard deviation. The error bars of $f_{\mathrm{iron}}^{\mathrm{planet}}$ cover the interval between the 16th and the 84th percentiles.}
\label{fig:firon_planet-star}
\end{figure}

We compare the iron mass fraction of the planets to the iron mass fraction of the protoplanetary disk, estimated from the host star composition. Based on planet interior models\cite{Dorn17}, we calculated the possible iron mass fraction of each planet ($f_{\mathrm{iron}}^{\mathrm{planet}}$) using only their mass and radius without incorporating any constraints from the host star composition. We considered two scenarios: i)  iron is present only in the core and ii) iron is present both in the core and mantle of each planet. Figure~\ref{fig:firon_planet-star} shows resulting relationship between $f_{\mathrm{iron}}^{\mathrm{planet}}$ and $f_{\mathrm{iron}}^{\mathrm{star}}$, which indicates a correlation between those quantities assuming either scenario. We again applied ODR and a $t$-test, finding that the correlation is statistically significant ($p$-value of $\sim$ 2$\times$10$^{-5}$). However, the planets span a wider range of iron-mass fraction than their host stars. The overall distribution of core-mass fraction (which can be related to the iron mass fraction) of rocky planets was shown to be wider than the overall distribution expected from the exoplanet host stars composition\cite{Plotnykov-20}. 

It has been suggested that the iron fraction in planets can be increased relative to the proto-stellar value if they formed close to rock-lines (regions where refractory material condensates/sublimates)\cite{Aguichine-20}. In this model, the enhancement in iron, however, is not high enough to explain the amount of iron in Mercury. This effect could explain the generally higher values of $f_{\mathrm{iron}}^{\mathrm{planet}}$ compared to $f_{\mathrm{iron}}^{\mathrm{star}}$. The trend we find in Figure~\ref{fig:firon_planet-star} suggests that if this effect operates, it must depend on the stellar iron-mass fraction, such that stars with higher iron fraction have a larger rock-line effect.

We identify a group of five planets (K2-38\,b, K2-106\,b, K2-229\,b, Kepler-107\,c, Kepler-406\,b) in Figure~\ref{fig:firon_planet-star} with higher content of iron than the rest of the planets. These five planets appear to be higher-mass analogues of Mercury, so we refer to them as 'super-Mercuries', a term previously proposed\cite{Marcus-10} by analogy with 'super-Earths' - planets with Earth-like compositions but higher masses. Several mechanisms of planet formation and evolution have been proposed to produce high-density and high $f_{\mathrm{iron}}^{\mathrm{planet}}$  super-Mercury planets\cite{Schulze-20}. The five super-Mercuries we identify have a wide range of masses, unlike the concentration around $\sim$ 5 $M_{\mathrm{\oplus}}$ predicted by studies of giant impacts\cite{Marcus-10}. We suggest that a giant impact alone is not responsible for the high density of super-Mercuries. Planet formation simulations that incorporate collisions are unable to produce the highest density super-Mercuries\cite{Scora-20}. There is a possible gap in the $f_{\mathrm{iron}}^{\mathrm{planet}}$ and $\rho_{\mathrm{Earth-like}}$  plane (Fig.~\ref{fig:firon_planet-star}) between super-Mercuries and super-Earths; we expect a continuous distribution from collisional stripping, given its stochastic nature.

Although we only find five super-Mercuries, they all orbit stars with  high $f_{\mathrm{iron}}^{\mathrm{star}}$ (meaning an overabundance of Fe relative to Mg and Si) and high iron abundance (Table S3), which is a proxy of the overall content of heavy elements in stars. The first trend may suggest that the mechanism responsible for the overabundance of iron in these planets is related to the composition of the proto-planetary disk. The second trend could imply a more efficient planet formation, leading to a formation of multiple planets and resulting in frequent collisions. We suggest that both iron enrichment\cite{Aguichine-20} and collision  mantle stripping\cite{Scora-20} may need to be invoked to  produce an iron enrichment in the general planet population and explain the presence of super-Mercuries.

Because the super-Mercuries may have an unusual origin and/or evolution, we investigated whether our findings still apply if we exclude them from our sample \cite{supl}. We find that the $f_{\mathrm{iron}}^{\mathrm{planet}}$ -- $f_{\mathrm{iron}}^{\mathrm{star}}$ correlation remains significant ($p$-value of $<$ 10$^{-4}$) for just the sample of super-Earths (Fig.~\ref{fig:firon_planet-star}), but shifts the slope to lower values (Tables S6 and S7). The difference in the slopes is, however, within the uncertainties: 6.3$\pm$1.2 against 4.3$\pm$0.8 for the case of $f_{\mathrm{core\&mantle}}^{\mathrm{planet}}$.  In addition, we found that the Fe/(Mg+Si) abundance ratio estimated for these planets correlates with the Fe/(Mg+Si) ratio of their host stars (Supplementary Text). Our results show a non 1-to-1 relationship with a slope greater than 1 (Supplementary Text).

All but one of the stars in our sample are members of the Milky Way's thin disk. The  exception is the ultra-short period  planet TOI-561 b, which orbits a metal-poor star in the thick disk; it was previously found to have an unusually low density\cite{Lacedelli-21}. Figs.~\ref{fig:density} and ~\ref{fig:firon_planet-star} show that the low density of this planet is consistent with the general trend  and dispersion we find for the entire  sample. Theory precits that planets orbiting around metal-poor thick disk and halo stars should have low iron mass fraction\cite{Santos-17}.

Several previous studies have searched for links between the composition of low-mass planets and their host stars. However, they were either based on single planetary systems\cite{Lillo-Box-20,Mortier-20}, on a small sample of planets\cite{Santos-15,Plotnykov-20,Schulze-20}, or on a statistical comparison of the overall populations\cite{Plotnykov-20}. None of those studies found a strong correlation due to these limitations.

The stellar abundances of major rock-forming elements, such as Fe, Mg, Si, are commonly used to infer the bulk compositions of rocky planets\cite{Dorn-15,Unterborn-16}, including in different stellar populations of the Milky Way\cite{Santos-17,Bitsch-20}. Our results provide support for the assumptions made in those studies. The observed correlation we find between $f_{\mathrm{iron}}^{\mathrm{planet}}$  and $f_{\mathrm{iron}}^{\mathrm{star}}$ has a slope that differs from 1 ($f_{\mathrm{iron}}^{\mathrm{planet}}$ is larger than $f_{\mathrm{iron}}^{\mathrm{star}}$); we interpret this as indicating that the composition of the protoplanetary disk (which vary with time and location) influences the resulting composition of planets in a non-linear fashion.

\newpage
\scriptsize

\textbf{Acknowledgments:} Based on data collected at Subaru Telescope (program ID S16B0178S; PI: V. Adibekyan), which is operated by the National Astronomical Observatory of Japan. We are honored and grateful for the opportunity of observing the Universe from Maunakea, which has the cultural, historical and natural significance in Hawaii. This research is also partly based on observations (program ID GN-217A-FT-20; PI: E. Jofr\'e) obtained at the international Gemini Observatory, a program of NSF’s NOIRLab, which is managed by the Association of Universities for Research in Astronomy (AURA) under a cooperative agreement with the National Science Foundation on behalf of the Gemini Observatory partnership: the National Science Foundation (United States), National Research Council (Canada), Agencia Nacional de Investigaci\'{o}n y Desarrollo (Chile), Ministerio de Ciencia, Tecnolog\'{i}a e Innovaci\'{o}n (Argentina), Minist\'{e}rio da Ci\^{e}ncia, Tecnologia, Inova\c{c}\~{o}es e Comunica\c{c}\~{o}es (Brazil), and Korea Astronomy and Space Science Institute (Republic of Korea).
This work has made use of data from the European Space Agency (ESA) mission (Gaia), processed by the Gaia Data Processing and Analysis Consortium (dpac). Funding for the dpac has been provided by national institutions, in particular the institutions participating in the Gaia Multilateral Agreement. This research has made use of the NASA Exoplanet Archive, which is operated by the California Institute of Technology, under contract with the National Aeronautics and Space Administration under the Exoplanet Exploration Program. This research has made use of the SIMBAD database, operated at CDS, Strasbourg, France.
V.A. thanks Akito Tajitsu for reducing the HDS spectra.
\textbf{Funding:} V.A., S.G.S., N.S., S.C.C.B., E.D.M., O.D.S.D., J.P.F. were supported by Funda\c{c}\~ao para a Ci\^encia e Tecnologia (FCT) through national funds and by FEDER through COMPETE2020 - Programa Operacional Competitividade e Internacionaliza\c{c}\~ao by these grants: UID/FIS/04434/2019; UIDB/04434/2020; UIDP/04434/2020; PTDC/FIS-AST/32113/2017 \& POCI-01-0145-FEDER-032113; PTDC/FIS-AST/28953/2017 \& POCI-01-0145-FEDER-028953. V.A., E.D.M, N.C.S., and S.G.S. acknowledge support from FCT through Investigador FCT contracts nr.  IF/00650/2015/CP1273/CT0001, IF/00849/2015/CP1273/CT0003, IF/00169/2012/CP0150/CT0002,   and IF/00028/2014/CP1215/CT0002, respectively, and POPH/FSE (EC) by FEDER funding through the program ``Programa Operacional de Factores de Competitividade - COMPETE''. C.D. acknowledges support from the Swiss National Science Foundation under grant PZ00P2\_174028, and the National Center for Competence in Research PlanetS supported by the SNSF. O.D.S.D. and  J.P.F. are supported by contracts (DL 57/2016/CP1364/CT0004 and DL57/2016/CP1364/CT0005, respectively) funded by FCT. E.M. acknowledges funding from the French National Research Agency (ANR) under contract number ANR-18-CE31-0019 (SPlaSH). B.B., was supported by the European Research Council (ERC Starting Grant 757448-PAMDORA).
\textbf{Author contributions:} V.A. led the data analysis and wrote the paper with contributions from C.D.. C.D. led the planetary interior analysis and S.G.S. performed the stellar parameter analysis. V.A. determined the composition of the stars with contributions from E.D.M.. N.C.S., B.B., C.M., B.M.T.B.S., and G.I. contributed to the discussion of the implications of the data. S.C.C.B., O.D.S.D., A.A.H, M.O. contributed to the general discussion of the results. C.D, S.G.S, N.C.S., G.I., C.M., E.D.M., M.K., Y.T., E.J., R.P., and E.M. gathered the spectroscopic observations. J.P.F. and P.F. performed to the statistical analysis. All authors discussed the results and commented on the manuscript.
\textbf{Competing interests: }We declared no competing interests.
\textbf{Data and materials availability:} The archival observations we used were obtained from the sources listed in Table S1. Our own observations are available in the observatory archives at \url{https://archive.gemini.edu/searchform} under program ID GN-2017A-FT-20 (Gemini) and \url{https://smoka.nao.ac.jp/objectSearch.jsp} by selecting Subaru HDS, then searching for “Kepler-20” and “Kepler-78”. Our final co-added spectra for all stars in
the sample are available on Figshare (30). The Python code we used to calculate the iron fractions is provided in Data S1. The output values for the properties of the stars and exoplanets are listed in Tables S3 and S4 respectively. The numerical values of the correlations we found using ODR are listed in Tables S6-S7.


\normalsize

\section*{ Materials/Methods, Supplementary Text, Tables, Figures, and References}
Materials and Methods\\
Supplementary Text \\
Figs. S1 -- S4\\
Tables S1 -- S8\\
Caption for Data S1\\
References \textit{(31-96)}

\newpage

\newpage

\section*{\hspace{2.0cm}Materials and Methods}
\label{Methods}

\vspace{0.5cm}
\subsection*{\underline{The sample}}

We started our sample selection from exoplanet.eu \cite{Schneider-11}. Out of 4330 confirmed planets (as of 2020 May 11) 364 were listed with masses below 10 $M_{\mathrm{\oplus}}$ orbiting F,G,K, type stars with effective temperatures ($T_{\mathrm{eff}}$)  between 4500 and 6500 K. We selected the 56 planets that had a precision in both mass and radius better than 30\%. To this sample we added two planetary systems  that were published after our search date: K2-38 (consisting of two low-mass planets)\cite{Toledo-Padron-20} and TOI-561 (consisting of one super-Earth and three mini-Neptunes)\cite{Lacedelli-21}. 

We then excluded all planets with mass estimations based on mass-radius empirical relations and planets with mass determination based on the Transit-timing variation (TTV) method, because planets with TTV- and RV-based mass determinations show different mass distributions\cite{Mills-17}. From the remaining 33 RV-detected exoplanets (Fig. 1) we selected 22 potentially rocky planets (Table S4) with raddi $<$ 2 R$_{\oplus}$ orbiting 21 stars.

\subsection*{\underline{Spectroscopic data}}

For the sample of 20 host stars we collected high-resolution ($\Re > 48~000$) optical spectra from seven spectrographs (Table~S\ref{tab:spectra}). Previous work has shown that the impact of using different instruments for spectroscopic analysis of host stars is very small\cite{Sousa-18, Adibekyan-20}. For K2-38 we used the results of the spectrosopic analysis based on the Echelle SPectrograph for Rocky Exoplanets and Stable Spectroscopic Observations (ESPRESSO, \citen{Pepe-21}) spectrum \cite{Toledo-Padron-20}.  One star (HD 80653) does not have an available high-resolution spectrum so was excluded from our sample.

The Echelle SpectroPolarimetric Device for the Observation of Stars (ESPaDOnS, \citen{Donati-03}) spectrum of HD 219134,  and the High Resolution Echelle Spectrometer (HIRES, \citen{Vogt-94}) of Kepler-107 and Kepler-406  are taken from published works \cite{Andreasen-17, Petigura-17}. Archival spectra of 11 stars (55~Cnc, EPIC~249893012, HD~213885, HD~3167, K2-106, K2-141, K2-216, K2-229, K2-265, K2-291, TOI-402) are taken from  High Accuracy Radial velocity Planet Searcher (HARPS, \citen{Mayor-03}) at the European Southern Observatory (ESO) 3.6 m Telescope, spectra of two stars (Kepler-10 and TOI-561) are taken from HARPS North (HARPS-N, \citen{Cosentino-12}) at the Spanish Observatorio del Roque de los Muchachos 3.57-m Telescopio Nazionale Galileo (TNG), and a spectrum of one star (WASP-47) is taken from the Fiber-fed Extended Range Optical Spectrograph (FEROS, \citen{Kaufer-99}) at the Max Planck Institute for Astronomy (MPG) and ESO 2.2 m Telescope. All the available individual spectra for each star were first corrected for their Dopler RV shifts and then were combined into a single spectrum with increased signal-to-noise ratio \cite{Sousa-18}. RV of the individual spectra were determined using \textsc{ares} v2 code\cite{Sousa-15}.

We obtained two consecutive spectra of Kepler-93 (2017 June 08 and 09) with the Gemini Remote Access to CFHT (Canada France Hawaii Telescope) ESPaDOnS Spectrograph (GRACES, \citen{Chene-14}) in the 1-fiber (object only) mode at the 8.1-m Gemini North telescope.  This mode provides a spectral resolution of $\Re \sim $  67~500 and spectral range of 420-1010 nm, characteristics that are appropriate to derive atmospheric parameters and elemental abundances. These spectra were reduced using the \textsc{opera} software\cite{Martioli-12} following published method\cite{Jofre-20}. The reduced spectra were then co-added into a single spectrum.

The observations of Kepler-20 and Kepler-78 were carried out with the High Dispersion Spectrograph (HDS, \citen{Noguchi-02}) on the 8.2m Subaru Telescope of the National Astronomical Observatory of Japan (NAOJ) on 2017 April 08. We considered  a slit size of 0.5 arcsecond to get a resolution of $\Re \sim $  72~000  and a wavelength Setup ”StdYc”. Two consecutive exposures were taken for each star. The reduction of the individual spectra was performed by Subaru staff, with Image Reduction and Analysis Facility (\textsc{iraf}, \citen{Tody-86}) echelle package\cite{Aoki-14}. The reduced spectra were then co-added into a single spectrum.

\subsection*{\underline{Stellar parameters and chemical abundances}}

Analysis of star-planet relations requires precise stellar parameters and elemental abundances for the host stars. In general, analysis of a population of exoplanets requires a homogeneous analysis of the host star spectra\cite{Ghezzi-10, Brugamyer-11, Santos-13}.

We determined the stellar atmospheric parameters ($T_{\mathrm{eff}}$, surface gravity $\log{g}$, microturbulent velocity $V_{\mathrm{tur}}$, and iron abundance relative to the solar value [Fe/H]) of the sample stars following published methods\cite{Sousa-14, Santos-13}. We use the equivalent widths (EW) of iron lines, measured in the combined spectra using \textsc{ares}, and we assume ionization and excitation equilibrium. We adopt a grid of Kurucz model atmospheres \cite{Kurucz-93} and the 2014 version of the radiative transfer code \textsc{moog} \cite{Sneden-73}. 

For K2-216, with a $T_{\mathrm{eff}}$ \ of $\sim$ 4500 K, the microturbulent velocity was close to zero, which we regard as unrealistically small. For this star we determined the microturbulence using an empirical calibration\cite{Adibekyan-12-kepler}. 

Stellar abundances of the elements were derived using the same tools and models and using a curve-of-growth analysis assuming local thermodynamic equilibrium. Although the EWs of the spectral lines were automatically measured with \textsc{ares}, we performed visual inspection of these measurements. For the abundances we followed previously published methods\cite{Adibekyan-12, Adibekyan-15}. The uncertainties of the abundances are calculated as a quadrature sum of the uncertainties from the EW measurements and those from the atmospheric parameters. The abundances of K2-38 are taken from published work which used the same methods and tools\cite{Toledo-Padron-20}.

The final stellar parameters and elemental abundances  (relative to the Sun) of the sample stars are presented in Table S\ref{tab:hosts2}.

We transformed the relative atmospheric abundances of Mg, Si, and Fe into absolute abundances\cite{Adibekyan-19} using the following solar reference values\cite{Asplund-09} and assuming the number of H atoms $\epsilon_{\mathrm{H}}$ is 10$^{12}$: $\log  \epsilon_{\mathrm{Mg}}$ = 7.6, $\log \epsilon_{\mathrm{Si}}$ = 7.51, and $\log \epsilon_{\mathrm{Fe}}$ = 7.5. We then used these absolute stellar abundances to estimate 

\begin{equation}
f_{\mathrm{iron}}^{\mathrm{star}} \equiv m_{\rm Fe} / (m_{\rm Fe} + m_{\rm MgSiO_{3}} + m_{\rm Mg_{2}SiO_{4}} + m_{\rm SiO_{2}}), \tag{S1}
\end{equation}

where $m_{\rm x} = N_{\rm x} \times \mu_{\rm x}$, $N_{\rm x}$  represents the number of atoms of each species X, and $\mu_{\rm x}$ is the mean molecular weight\cite{Santos-15, Santos-17}. The $N_{\rm x}$ values are computed relative to hydrogen.

When $1 < N_{\rm Mg}/N_{\rm Si} < 2$, which is the case for most of the stars, Mg and Si are assumed to be  equally distributed between pyroxene (MgSiO$_{3}$) and olivine (Mg$_{2}$SiO$_{4}$). The number of these compounds can then be calculated as: $N_{\rm MgSiO_{3}}$ = 2$N_{\rm Si}$ - $N_{\rm Mg}$ and $N_{\rm Mg_{2}SiO_{4}}$ = $N_{\rm Mg}$ - $N_{\rm Si}$. In case of $N_{\rm Mg}/N_{\rm Si} < 1$, we assume magnesium is incorporated only into pyroxene, leaving the remaining silicon to form mostly SiO$_{2}$. The number of these compounds is then: $N_{\rm MgSiO_{3}}$ = $N_{\rm Mg}$ and $N_{\rm SiO_{2}}$ =  $N_{\rm Si}$ - $N_{\rm Mg}$. No star in the sample has $N_{\rm Mg}/N_{\rm Si} > 2$.

The uncertainties in the iron mass fraction have been determined by using Monte Carlo simulations. We randomly drew 10$^{5}$ values of abundances cosidering a Gaussian distribution centered on the derived abundance with a standard deviation corresponding to the derived uncertainties of the abundances (as listed in Table S\ref{tab:hosts2}). The distribution of the resulting values of $f_{\mathrm{iron}}^{\mathrm{star}}$ allowed us to derive the one-sigma uncertainity which are listed in Table S\ref{tab:hosts2}.

The luminosity of the planet host stars was calculated by using spectroscopic effective temperature, V-bamd magnitude (Table S\ref{tab:hosts1}), Gaia DR2 parallax\cite{Gaia-18}, and bolometric correction\cite{Flower-96}. These parameters are listed in Table  S\ref{tab:hosts1}.

\subsection*{\underline{Physical properties of the rocky planets}}

The planetary parameters were taken from exoplanet.eu\cite{Schneider-11}. We computed the bulk density, $\rho$, of the planets based on their mass and radius. Because planets with the same composition but with different masses would have different bulk densities, we scaled the densities to the density of a planet with Earth-like composition\cite{Dorn17} for a given mass - $\rho_{\mathrm{Earth-like}}$. 

Given only the planet's mass and radius, we estimated their possible iron fraction: 

\begin{equation}
f_{\mathrm{iron}}^{\mathrm{planet}} \equiv (M_{\rm Fe, mantle}+M_{\rm core})/M, \tag{S2}
\end{equation}

where $M_{\rm Fe, mantle}$, $M_{\rm core}$ are the masses of iron in mantle and core, respectively and $M$ is the total mass of the planet. In addition, we also estimated the possible Fe/(Mg+Si) abundance ratio of the planets using an interior model \cite{agol2020refining} and a characterization scheme \cite{Dorn17} that employs a Markov chain Monte Carlo (McMC) method. 
For the planet interiors, we assume a pure iron core and a silicate mantle, and neglect any volatile layers. The interior model uses self-consistent thermodynamics in the core and mantle. For the core, we use the equations of state for hexagonal close packed iron\cite{Hakim18} and for the silicate mantle we use equation-of-state model \cite{sotin2007mass}. We assume an adiabatic temperature profile within core and mantle, with adiabatic gradients computed from the equations of states. 

When estimating $f_{\mathrm{iron}}^{\mathrm{planet}}$, we test two scenarios: First, we only allow the core to vary in size and fix the iron content of the mantle at zero (Fig. 3A). In a second scenario, we also allow the mantle composition to vary in iron content (Fig. 3B). This results in $\sim$10\% higher $f_{\mathrm{iron}}^{\mathrm{planet}}$, since high amounts of iron can be in part compensated by high amounts of oxygen to produce the same mass and radius. Average estimated Mg-numbers (Mg/(Mg+Fe)) of the mantles range from 0.45 to 0.9 with large uncertainties of up to 60\%. This is consistent with the long-range migration scenario in which the planets formed in a highly oxidizing environment, which enabled the iron to remain in the mantle\cite{Monteux-18}. The model values of FeO/Fe indicate the oxidation states, corresponding to average oxygen fugacities from -3 to -1.7$\Delta \mathrm{IW}$ which is similar to the oxidation state of Earth and small bodies, both in the Solar System and accreted by white dwarfs \cite{Doyle-19}.

Here, oxygen fugacity is defined relative to the Iron-W\"ustite equilibrium reaction $\mathbf{\mathrm{Fe}{+}0.5\mathrm{O}_2{=}\mathrm{FeO}}$ (W\"ustite) such that:

\begin{equation}
\Delta IW = 2 log(x^{rock}_{FeO} / x^{metal}_{Fe}) + 2 log(\gamma^{rock}_{FeO} / \gamma^{metal}_{Fe}), \tag{S3}
\end{equation}

where $x_{i}^{k}$ are mole fractions of the species $i$ in phase $k$, and $\gamma_{i}^{k}$ are activity coefficients for the species. This equation was introduced as an intrinsic oxygen fugacity of a planet\cite{Doyle-19}. The uncertain activity coefficients are set to unity such that the second term of the right-hand side of the equation vanishes.

In our analysis we neglected light elements in the core, since their addition has little effects on the total radius. For example, we find that for a fixed Fe/Si bulk ratio, the addition of light alloys like Fe$_{0.95}$O in the core only changes the radius by ~0.2\%,  using light alloy equations of state\cite{Hakim18}.

The planet structure model we adopt does not include an atmospheric layer. However, the atmospheric build-up on planets that lost their H$_{2}$-dominated envelopes is very limited, especially when planets are at close-in orbits\cite{Kite-20} which is the case for all the planets in our sample. If the planets of our sample have Earth-like atmospheres, then neglecting them would have negligible impact on our results. As a limiting case we consider the effect if the planets are assumed to have atmospheres as large as Venus's. The observations of the transit of Venus in 2012 in the visible (at 450nm) indicated a $\sim$ 80km atmosphere\cite{Reale-15}, which is ~1.3\% of its radius. This value is, however, still a few times smaller than the mean uncertainty of the radius of the exoplanets in our sample. Thus we conclude that neglecting the potential presence of atmospheres should not have a strong impact on our results and conclusions. Our sample includes a few low density planets that have $f_{\mathrm{iron}}^{\mathrm{planet}}$ smaller than $f_{\mathrm{iron}}^{\mathrm{star}}$, which could be partially explained by the presence of volatile outer layers\cite{Schulze-20}.

\begin{figure}
\begin{center}
\includegraphics[width=1\linewidth]{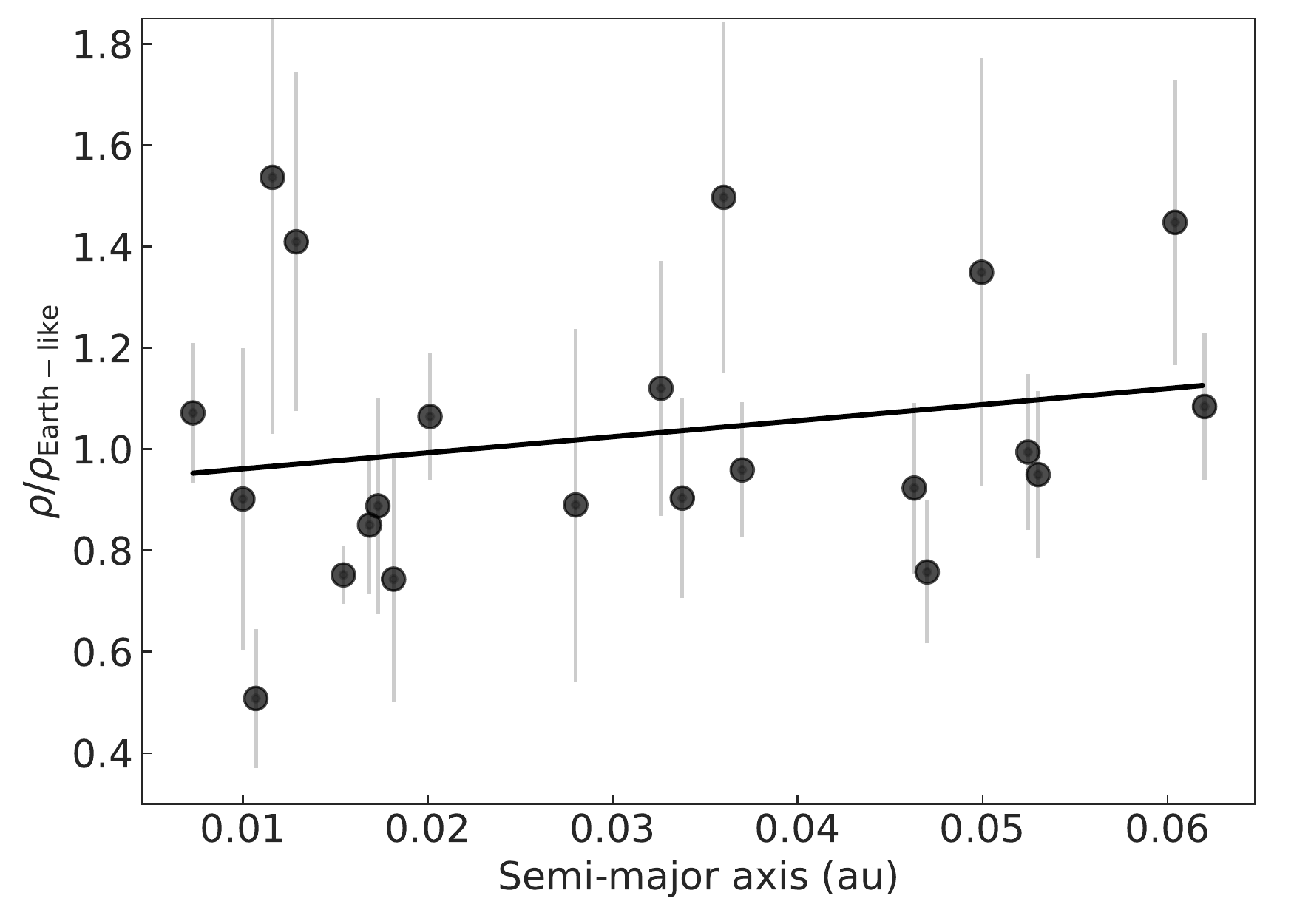}
\end{center}
\caption{\textbf{Fig. S1. Densities of rocky planets as a function of orbital distances.} The black solid line represent the results of the Ordinary Least Squares linear regression as implemented in \textsc{statsmodels}\cite{seabold2010statsmodels}. 
We then used the values of the slope and the associated uncertainty (3.2$\pm$3.3) to assess the significance of the correlation by applying a $t$-test. Based on the P($t$-stat) = 0.35 value we conclude that the correlation is not statistically significant. The error bars of normalized densities show one standard deviation. The semi-major axes are in astronomical units (au).}
\label{fig:S1}
\end{figure}

\begin{figure}
\begin{center}
\includegraphics[width=1\linewidth]{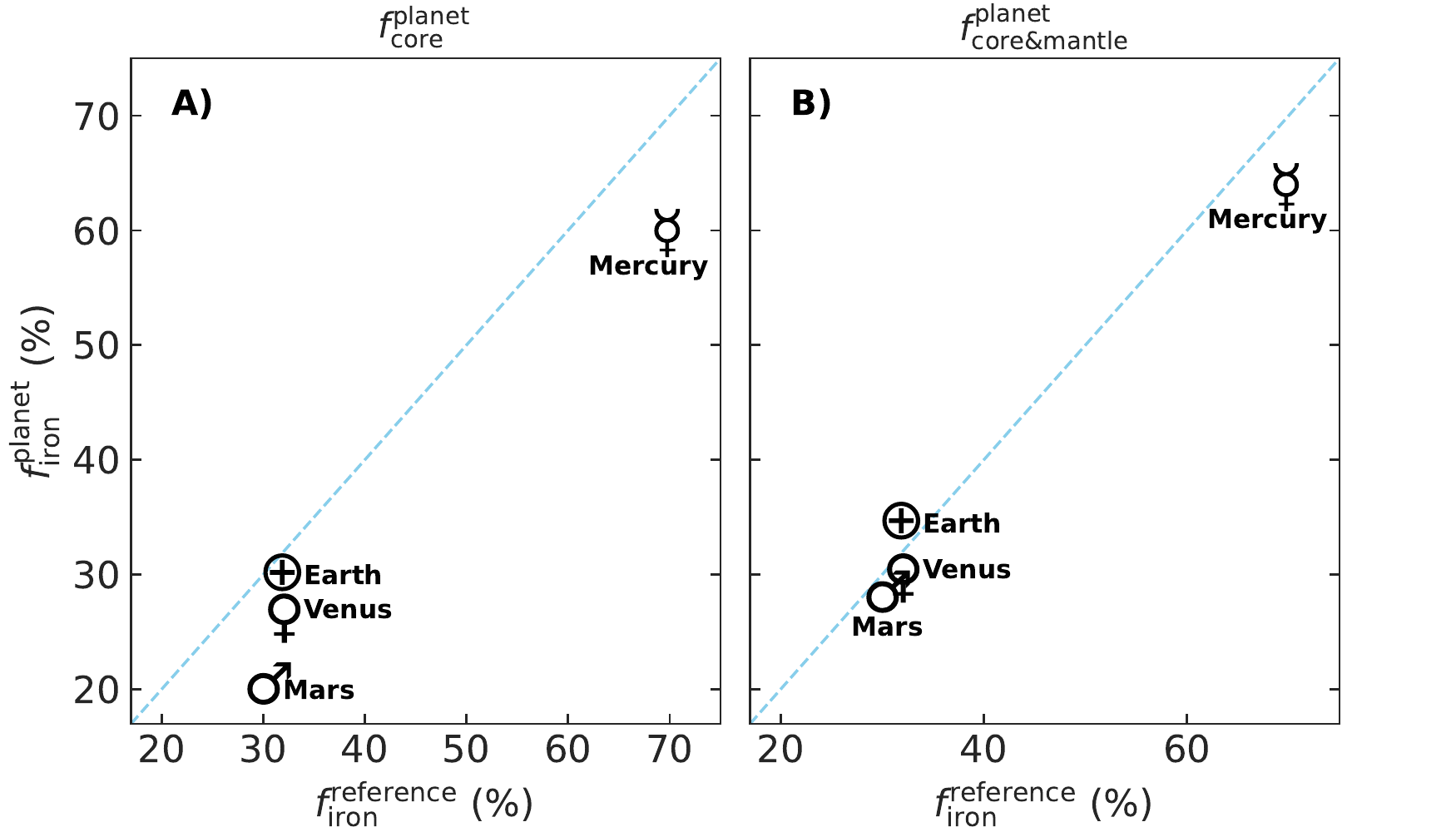}
\end{center}
\caption{\textbf{Fig. S2. Iron content in the rocky planets of our Solar System.} Iron mass fraction of the planets $f_{\mathrm{iron}}^{\mathrm{planet}}$ as estimated by their mass and radius and using an interior model\cite{Dorn17} compared to the measured values $f_{\mathrm{iron}}^{\mathrm{reference}}$ \cite{Hauck-13, Aitta-12, McDonough-03, Khan-18}. Estimates of $f_{\mathrm{iron}}^{\mathrm{planet}}$ are based on the assumption that: (A) all iron resides in the core only ($f_{\mathrm{core}}^{\mathrm{planet}}$) or (B) iron is assumed to be present in both mantle and core ($f_{\mathrm{core\&mantle}}^{\mathrm{planet}}$). The identity lines are shown as sky-blue dashed lines.}
\label{fig:S2}
\end{figure}

For comparison, using the same planet interior model we also determined the iron mass fraction of the Solar System rocky planets. The following reference values have been adopted: $f_{\mathrm{iron}}^{\mathrm{reference}}$ of $\sim$ 70\% for Mercury\cite{Hauck-13}, $f_{\mathrm{iron}}^{\mathrm{reference}}$ of $\sim$ 32\% for Venus\cite{Aitta-12}, $f_{\mathrm{iron}}^{\mathrm{reference}}$ of $\sim$ 32\% for Earth\cite{McDonough-03}, and $f_{\mathrm{iron}}^{\mathrm{reference}}$ of $\sim$ 30\% for Mars\cite{Khan-18}. Figure S\ref{fig:S2} compares our model values to the measured reference values; we find an average difference of $\sim$7$\pm$4 \% and $\sim$2$\pm$4 \% 
when we allow iron to be only in the core (Fig. S\ref{fig:S1}A) and to be both in the core and mantle (Fig. S\ref{fig:S1}B), respectively.

The $T_{\mathrm{eq}}$ of the exoplanets were computed using the stellar luminosity and orbital distances of the planets assuming zero Bond albedo\cite{Wang-18}. The  semi-major axes for K2-141~b and Kepler-406~b were not available in exoplanet.eu so were extracted from published works \cite{Barragan-18, Butler-06}.

The parameters of the rocky planets of our sample plus those of the Solar System are presented in Table S\ref{tab:planets}. Our planet sample has much higher equilibrium temperatures than the Solar System planets. It is thus possible that the planets of our sample have been affected by processes that did not affect the Solar System planets.

\subsection*{\underline{Statistical significance and robustness of the results}}

We performed an orthogonal distance regression (ODR) as implemented in \textsc{scipy}\cite{Virtanen-20} to quantify the observed $\rho / \rho_{\mathrm{Earth-like}}$ -- $f_{\mathrm{iron}}^{\mathrm{star}}$ and $f_{\mathrm{iron}}^{\mathrm{planet}}$ -- $f_{\mathrm{iron}}^{\mathrm{star}}$ correlations. We use the inverse variance on both quantities to weigh each data point. The results of the ODR analysis are summarized in Table S\ref{tab:statistics}, which lists the intercept and slope of the regression lines together with the associated uncertainties, as well as the residual variance which is a measure of goodness-of-fit. We then used the values of the slopes and the associated uncertainties to assess the significance of the correlation. We used  $t$-statistics (as implemented in \textsc{scipy}) to test the null hypothesis  that the data can be represented with a model that has  a null slope. The resulting two-tailed p-values (P($t$-stat)) are presented in Table S\ref{tab:statistics}. We conclude that the correlations are significant.

\begin{figure}
\begin{center}
\includegraphics[width=1\linewidth]{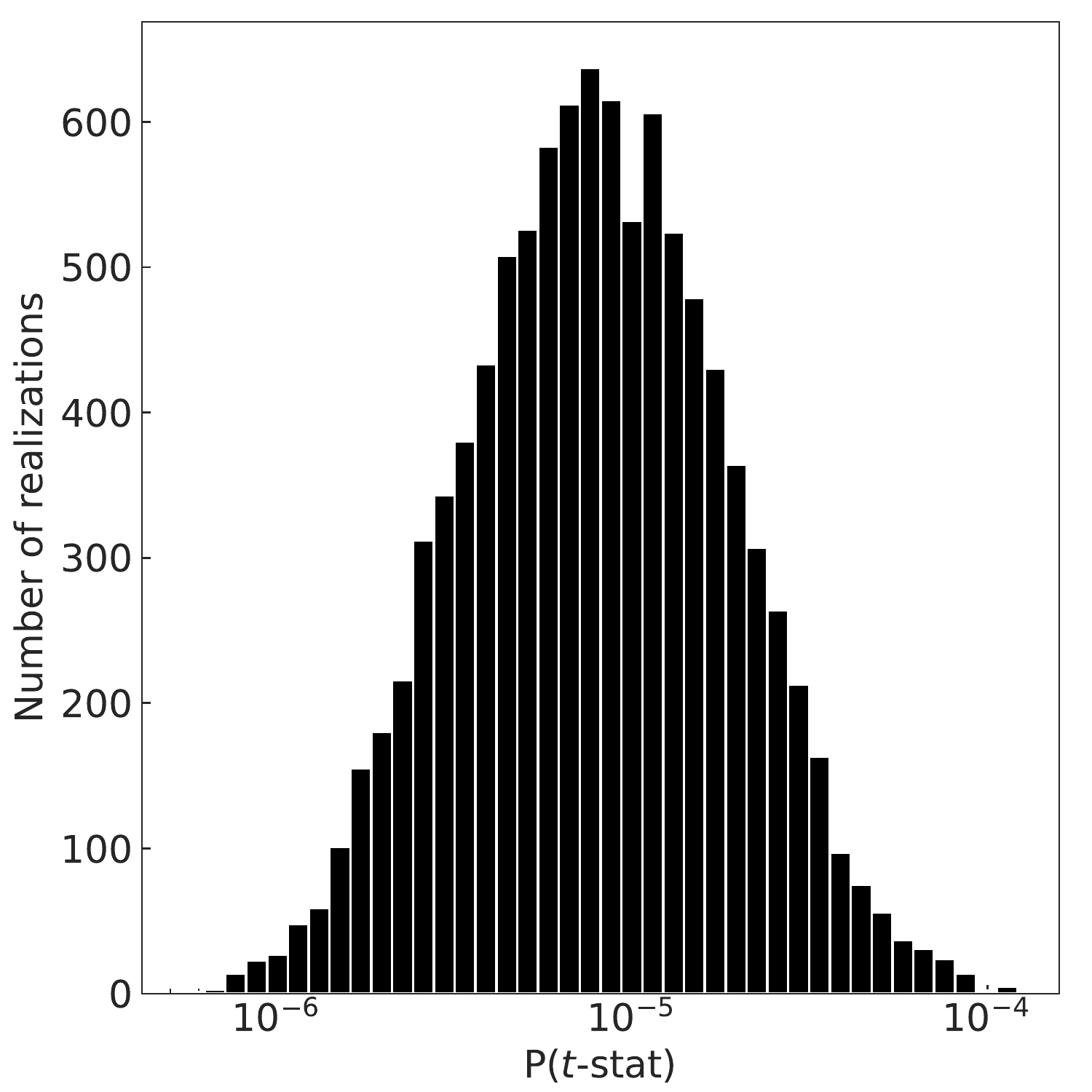}
\end{center}
\caption{\textbf{Fig. S3. Distribution of the P($t$-stat) for the $\rho / \rho_{\mathrm{Earth-like}}$ vs $f_{\mathrm{iron}}^{\mathrm{star}}$  correlation.} Different combinations of planetary masses and radii measurements listed in Table~S\ref{tab:planets_2} are used to estimate the $\rho / \rho_{\mathrm{Earth-like}}$. The ODR and a $t$-test was applied to quantify the dependence of $\rho / \rho_{\mathrm{Earth-like}}$ on $f_{\mathrm{iron}}^{\mathrm{star}}$ for each combination. Based on the small values of P($t$-stat) we conclude that our results are robust to the various sources of planetary parameters.}
\label{fig:p_values}
\end{figure}

Since the planetary parameters used in this work are not determined homogeneously, but taken from the exoplanets.eu database we performed another test to explore the robustness of our results to the heterogeneity in the planet data. From the NASA Exoplanet Archive\cite{NEA} we selected up to three most recent published planetary parameters which satisfy the 30\% relative uncertainty threshold we adopted above (Table S\ref{tab:planets_2}). For the majority of the planets only one or two measurements were available in the  database, while a few planets (for example, Kepler-78 b and 55 Cnc e) had more measurements. For 55 Cnc e, we ignored radius estimates above 2R$_{\oplus}$, because in these cases the planet would lie above the radius gap and would not be included in our sample.  We then considered all the possible combinations of the planetary parameters (given the available planetary parameters for each system) and performed ODR regressions to quantify the $\rho / \rho_{\mathrm{Earth-like}}$ vs $f_{\mathrm{iron}}^{\mathrm{star}}$  correlation. This test showed that the P($t$-stat) values are smaller than 1$\times$10$^{-4}$ with the mean value being 1$\times$10$^{-5}$ (Fig. S\ref{fig:p_values}). The results of this test confirm the statistically significance of our results and show that they are robust to the various sources of planetary parameters.

These tests show that the correlation between the composition of the stars and the properties of the rocky planets is statistically significant. However, they do not establish that the observed correlation is of astrophysical origin. Determination of both the stellar composition and planetary properties (mass, radius, and associated density and $f_{\mathrm{iron}}^{\mathrm{planet}}$)  depend on the stellar atmospheric parameters which might, in principle, produce a non-astrophysical correlation between these quantities.

The planetary masses and radii, determination of which directly depends on the stellar masses and radii, were determined independently in previous studies. The stellar parameters used in those studies have been determined using different methods, and often different from the method that we adopted in the current work. A heterogeneous planet properties determination will most probably introduce a scatter rather than a correlation. However, we established above that the results are robust to the heterogeneity in the planet data.

Individual stellar abundances of Mg, Si, and Fe are more sensitive to the variation of stellar parameters than their ratios, because of their similar sensitivity to the stellar parameters\cite{Adibekyan-12}. If the sensitivity to the errors of stellar parameters was the source of the observed correlation then one would observe a correlation between planetary properties and the abundances of any element, which is not the case: $\rho / \rho_{\mathrm{Earth-like}}$  and $f_{\mathrm{iron}}^{\mathrm{planet}}$ do not correlate with the individual stellar abundances (Table S\ref{tab:statistics_tests}). Given the dependency of the individual abundances on each of the stellar parameters\cite{Adibekyan-12}, and the values of $f_{\mathrm{iron}}^{\mathrm{planet}}$ and $f_{\mathrm{iron}}^{\mathrm{star}}$  it would be difficult to introduce a positive correlation between these two parameters.

Finally, the range of stellar parameters in our sample is larger than the typical uncertainties of these parameters. The sample includes stars with very different stellar atmospheric parameters which have similar $f_{\mathrm{iron}}^{\mathrm{star}}$ and/or planetary properties, suggesting that the compositional link between the stars and their planets is not due to the covariance with the stellar atmospheric parameters.

\vspace{2cm}
\section*{\hspace{2.0cm}Supplementary Text}

\vspace{0.5cm}
\subsection*{\underline{Super-Earths and super-Mercuries}}

Figure~3 shows the planets of our sample can be separated into super-Mercuries and super-Earths, where super-Mercuries are the planets with the highest $f_{\mathrm{iron}}^{\mathrm{planet}}$. For the sample of super-Earths we performed the same statistical test as in the previous section. The results of the test are presented in Table S\ref{tab:statistics_superEarths} and show that the relation between $f_{\mathrm{iron}}^{\mathrm{planet}}$ and $f_{\mathrm{iron}}^{\mathrm{star}}$ is statistically significant (P($t$-stat) values $<$ 1$\times$10$^{-4}$). The residual variance of the fits for this sub-sample are  smaller than the corresponding values obtained for the full sample (see Table S\ref{tab:statistics}). This indicates that the model is a better match to the data.

\begin{figure}
\begin{center}
\includegraphics[width=1\linewidth]{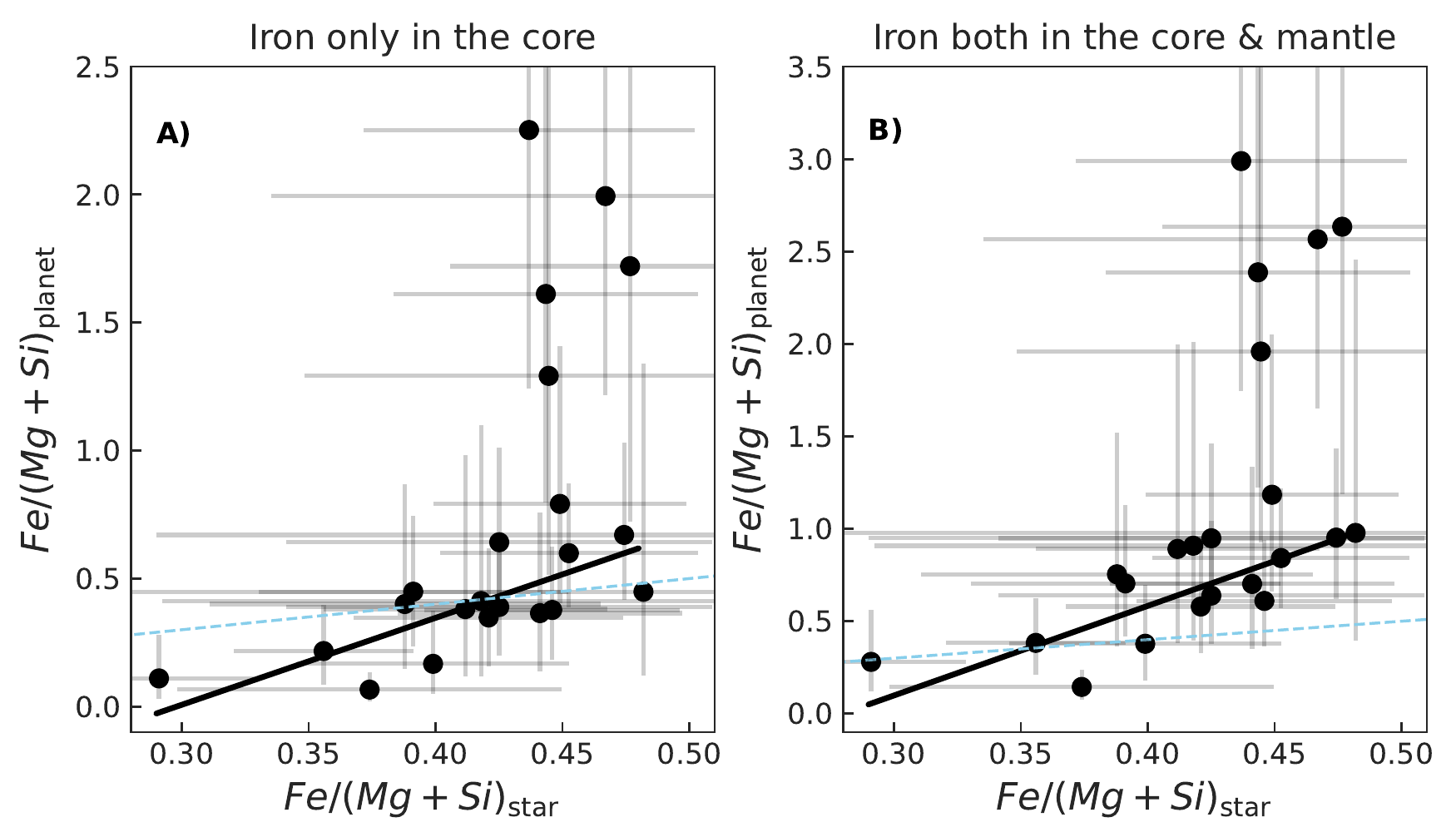}
\end{center}
\caption{\textbf{Fig. S4. Abundance ratios in planets and their hosts.} Fe/(Mg+Si) abundance ratio of the protoplanetary disc estimated from the host star composition as a function of the Fe/(Mg+Si) ratio of the planets as estimated from their mass and radius and using an interior model\cite{Dorn17}. The estimates for planets are based on the assumption that: (A) all iron resides in the core only or (B) iron is assumed to be present in both mantle and core (right panel). The black solid lines represent the results of the ODR linear regression for super-Earths only i.e. excluding the five possible super-Mercuries. Results are presented in Methods. The identity lines are shown as light-blue dashed lines. The error bars of Fe/(Mg+Si) star show one standard deviation. The error bars of Fe/(Mg+Si) planet cover the interval between the 16th and the 84th percentiles.}
\label{fig:fig_Mg_fe_si_star_planet}
\end{figure}

In Figure~S\ref{fig:fig_Mg_fe_si_star_planet} we show the dependence of Fe/(Mg+Si) abundance ratio of the planets on the same abundance ratio derived for the host stars. The super-Earths and super-Mercuries are separated in this plot. While the sample of super-Mercuries is small and it is not possible to conclude whether a correlation exist for these planets alone, the sample of super-Earths has a strong correlation with a slope $>$ 3.

For the Super-Earths only, we performed an orthogonal distance regression analysis to quantify the Fe/(Mg+Si)$_{\mathrm{star}}$ -- Fe/(Mg+Si)$_{\mathrm{planet}}$ relation considering uncertainties of both variables. The results are slightly different for the two assumptions we made about the iron content in the planets: 
\begin{equation}
 Fe/(Mg+Si)_{\mathrm{core}}^{\mathrm{planet}}  = -1.01(\pm0.25)  + 3.39(\pm0.64) \times Fe/(Mg+Si)_{\mathrm{star}} \tag{S4}
\end{equation}
\begin{equation}
Fe/(Mg+Si)_{\mathrm{core\&mantle}}^{\mathrm{planet}}  = -1.35(\pm0.36)  + 4.84(\pm0.92) \times Fe/(Mg+Si)_{\mathrm{star}}, \tag{S5}\\
\end{equation}

where $Fe/(Mg+Si)_{\mathrm{core}}^{\mathrm{planet}}$ and $Fe/(Mg+Si)_{\mathrm{core\&mantle}}^{\mathrm{planet}}$ correspond to the cases when iron is allowed to be only in the core, and in the core and mantle  of planets, respectively. In both cases, however, the relation has a slope different from 1.

As in the previous section, we assessed the significance of these relations by calculating the P($t$-stat), which are $\sim$ 1$\times$10$^{-4}$ for both cases. The residual variance of the fits are similar too, being $\sim$0.18.

\newpage

\begin{table}[ht]
\caption{\label{tab:spectra} \textbf{Table S1. Sources of the host star spectra.} The second column lists the International Celestial Reference System (ICRS) coodrinates of the stars at the J2000 epoch. The fourth and fifth  columns lists the name of the Principal Investigator (PI) and the observing proposal identifier (ID). The last column indicates the sources of the spectra. }
\scriptsize
{\renewcommand{\arraystretch}{1.3}
\begin{tabular}{llllll}
\hline
     Star & ICRS coordinates, J2000 &  Spectrograph &  PI & ID & Source\\
\hline
55Cnc  &  133.14921 +28.33082  &  HARPS  &  Triaud  &  288.C-5010(A)  &  ESO archive$^{*}$ \\ 
EPIC249893012  &  228.24817 -16.72450  &  HARPS  &  Armstrong  &  1102.C-0249(A)  &  ESO archive$^{*}$ \\ 
EPIC249893012  &  228.24817 -16.72450  &  HARPS  &  Gandolfi  &  0101.C-0829(A)  &  ESO archive$^{*}$ \\ 
EPIC249893012  &  228.24817 -16.72450  &  HARPS  &  Gandolfi  &  1102.C-0923(A)  &  ESO archive$^{*}$ \\ 
HD213885  &  338.98465 -59.86447  &  HARPS  &  Diaz  &  0102.C-0525(A)  &  ESO archive$^{*}$ \\ 
HD213885  &  338.98465 -59.86447  &  HARPS  &  Armstrong  &  1102.C-0249(A)  &  ESO archive$^{*}$ \\ 
HD213885  &  338.98465 -59.86447  &  HARPS  &  Jordan  &  0101.C-0510(C)  &  ESO archive$^{*}$ \\ 
HD 219134  &  348.32073 +57.16835  &  ESPaDOnS  &    &    &  \cite{Andreasen-17} \\ 
HD3167  &  \,\, \,8.73968   \,\,  +4.38147  &  HARPS  &  Gandolfi  &  098.C-0860(A)  &  ESO archive$^{*}$ \\ 
HD3167  &  \,\, \,8.73968   \,\,  +4.38147  &  HARPS  &  Gandolfi  &  097.C-0948(A)  &  ESO archive$^{*}$ \\ 
K2-106  &  \,\, 13.07976 +10.79470  &  HARPS  &  Gandolfi  &  098.C-0860(A)  &  ESO archive$^{*}$ \\ 
K2-141  &  350.91655  \,\,-1.18929  &  HARPS  &  Gandolfi  &  099.C-0491(A)  &  ESO archive$^{*}$ \\ 
K2-216  &  \,\, 11.48024  \,\,+6.34698  &  HARPS  &  Gandolfi  &  0100.C-0808(A)  &  ESO archive$^{*}$ \\ 
K2-216  &  \,\, 11.48024  \,\,+6.34698  &  HARPS  &  Gandolfi  &  098.C-0860(A)  &  ESO archive$^{*}$ \\ 
K2-216  &  \,\, 11.48024  \,\,+6.34698  &  HARPS  &  Gandolfi  &  099.C-0491(A)  &  ESO archive$^{*}$ \\ 
K2-229  &  186.87327  -6.72188  &  HARPS  &  Santerne  &  198.C-0169(A)  &  ESO archive$^{*}$ \\ 
K2-265  &  342.03152 -14.49468  &  HARPS  &  Santerne  &  198.C-0169(A)  &  ESO archive$^{*}$ \\ 
K2-291  &  \,\, 76.44580 +21.54862  &  HARPS  &  Jordan  &  0101.C-0510(C)  &  ESO archive$^{*}$ \\ 
Kepler-10  &  285.67942 +50.24131  &  HARPS-N  &  Pepe  &  GTO$^{\ddag}$  &  TNG archive$^{\dag}$ \\ 
Kepler-107  &  297.02822 +48.20860  &  HIRES  &    &    &  \cite{Petigura-17} \\ 
Kepler-20  &  287.69801 +42.33869  &  HDS  &  Adibekyan  &  S16B0178S  &  This work \\ 
Kepler-406  &  291.84807 +44.96826  &  HIRES  &    &    &  \cite{Petigura-17} \\ 
Kepler-78  &  293.74172 +44.44832  &  HDS  &  Adibekyan  &  S16B0178S  &  This work \\ 
Kepler-93  &  291.41829 +38.67234  &  GRACES  &  Jofre  &  GN-217A-FT-20  &  This work \\ 
TOI-402  &  \,\, 36.86824 -27.63521  &  HARPS  &  Diaz  &  198.C-0836(A)  &  ESO archive$^{*}$ \\ 
TOI-402  &  \,\, 36.86824 -27.63521  &  HARPS  &  Mayor  &  072.C-0488(E)  &  ESO archive$^{*}$ \\ 
TOI-402  &  \,\, 36.86824 -27.63521  &  HARPS  &  Udry  &  183.C-0972(A)  &  ESO archive$^{*}$ \\ 
TOI-402  &  \,\, 36.86824 -27.63521  &  HARPS  &  Udry  &  192.C-0852(A)  &  ESO archive$^{*}$ \\ 
TOI-402  &  \,\, 36.86824 -27.63521  &  HARPS  &  Udry  &  196.C-1006(A)  &  ESO archive$^{*}$ \\ 
TOI-561  &  148.18562 \,\, +6.21637  &  HARPS-N  &  Piotto  &  A40TAC\_23  &  TNG archive$^{\dag}$ \\ 
WASP-47  &  331.20303 -12.01889  &  FEROS  &  Faedi  &  089.C-0471(A)  &  ESO archive$^{*}$ \\ 
WASP-47  &  331.20303 -12.01889  &  FEROS  &  Sousa  &  089.C-0444(A)  &  ESO archive$^{*}$ \\
 \\
\hline
\end{tabular}
}
\\
{\raggedright * \url{http://archive.eso.org/eso/eso_archive_main.html} \par}
{\raggedright \dag \url{http://archives.ia2.inaf.it/tng/} \par}
{\raggedright \ddag Guaranteed Time Observations. \par}
\end{table}

\begin{table}[ht]
\caption{\label{tab:hosts1} \textbf{Table S2. Properties of the host stars.} The seventh column (References) lists the sources for the $V$-band magnitudes. The bolometric correction (column 4), the $V$-band absolute magnitudes (column 5), and the luminocities (column 6) are derived in this work.}
\scriptsize
{\renewcommand{\arraystretch}{1.9}
\setlength{\tabcolsep}{2.7pt}
\begin{tabular}{llllllll}
\hline
     Star & ICRS coordinates, J2000 &  Parallax (mas) & $V$-band (mag) & $BC_{\mathrm{v}}$ (mag) & $M_{\mathrm{v}}$-band  (mag) & Luminocity ($L_{\mathrm{\odot}}$) & References\\
\hline
K2-38  & 240.03358 -23.18926 & 5.16 & 11.45 & -0.09 & 4.90 & 0.93 &  \cite{Hog-00} \\ 
K2-106  & \,\, 13.07976 +10.79470 & 4.06 & 12.16 & -0.13 & 5.14 & 0.77 &  \cite{Hog-00} \\ 
K2-229  & 186.87327  -6.72188 & 9.72 & 11.10 & -0.23 & 5.92 & 0.41 &  \cite{Hog-00} \\ 
Kepler-107  & 297.02822 +48.20860 & 1.87 & 12.70 & -0.05 & 3.82 & 2.43 &  \cite{Bonomo-19} \\ 
Kepler-406  & 291.84807 +44.96826 & 2.72 & 12.52 & -0.11 & 4.65 & 1.19 &  \cite{Schneider-11} \\ 
K2-291  & \,\, 76.44580 +21.54862 & 11.07 & 10.11 & -0.13 & 5.23 & 0.71 &  \cite{Hog-00} \\ 
55 Cnc  & 133.14921 +28.33082 & 79.43 & 6.03 & -0.18 & 5.21 & 0.76 &  \cite{Hog-00} \\ 
EPIC 249893012  & 228.24817 -16.72450 & 3.08 & 11.53 & -0.13 & 4.02 & 2.16 &  \cite{Hog-00} \\ 
HD 213885  & 338.98465 -59.86447 & 20.77 & 7.94 & -0.06 & 4.53 & 1.28 &  \cite{Hog-00} \\ 
HD 219134  & 348.32073 +57.16835 & 153.08 & 5.67 & -0.41 & 6.49 & 0.29 &  \cite{Hog-00} \\ 
HD 3167  & \,\, \,8.73968   \,\,  +4.38147 & 21.12 & 9.03 & -0.19 & 5.38 & 0.66 &  \cite{Hog-00} \\ 
K2-141  & 350.91655  \,\,-1.18929 & 16.13 & 11.53 & -0.61 & 7.49 & 0.14 &  \cite{Hog-00} \\ 
K2-216  & \,\, 11.48024  \,\,+6.34698 & 8.63 & 12.48 & -0.60 & 7.16 & 0.19 &  \cite{Zacharias-13} \\ 
K2-265  & 342.03152 -14.49468 & 7.18 & 11.27 & -0.15 & 5.47 & 0.58 &  \cite{Hog-00} \\ 
Kepler-10  & 285.67942 +50.24131 & 5.36 & 11.16 & -0.10 & 4.56 & 1.29 &  \cite{Dumusque-14} \\ 
Kepler-20  & 287.69801 +42.33869 & 3.51 & 12.51 & -0.14 & 5.24 & 0.71 &  \cite{Lasker-08} \\ 
Kepler-78  & 293.74172 +44.44832 & 8.01 & 11.85 & -0.29 & 6.24 & 0.33 &  \cite{Hog-00} \\ 
Kepler-93  & 291.41829 +38.67234 & 10.40 & 10.08 & -0.11 & 5.09 & 0.80 &  \cite{Hog-00} \\ 
TOI-402  & \,\, 36.86824 -27.63521 & 22.28 & 9.18 & -0.27 & 5.83 & 0.47 &  \cite{Hog-00} \\ 
WASP-47  & 331.20303 -12.01889 & 3.75 & 11.99 & -0.12 & 4.86 & 1.00 &  \cite{Zacharias-13} \\ 
TOI-561  & 148.18562 \,\, +6.21637 & 11.63 & 10.33 & -0.19 & 5.58 & 0.55 &  \cite{Hog-00} \\ 
\hline
\end{tabular}
}
\\
\end{table}

\begin{table}[ht]
\caption{\label{tab:hosts2} \textbf{Table S3. Atmospheric parameters and compositions of the host stars.} The atmospheric parameters and elemental abundances of K2-38 are taken from published work\cite{Toledo-Padron-20}. The properties of the rest of the exoplanet hosts stars are derived from our spectral analysis as described in Methods. The abundances of Fe, Mg, and Si are relative to the Sun. The solar atmospheric parameters and absolute abundances are taken from published works\cite{Adibekyan-16, Asplund-09}.}
\scriptsize
{\renewcommand{\arraystretch}{1.9}
\begin{tabular}{llllllllll}
\hline
     Star  & $T_{\mathrm{eff}}$ (K) & $\log{g}$ (dex) & $V_{\mathrm{tur}}$ (km/s)  &  [Fe/H] & [Mg/H] &  [Si/H] &  $f_{\mathrm{iron}}^{\mathrm{star}}$ (\%) \\
\hline
 K2-38 & 5731$\pm$66$^{*}$ & 4.38$\pm$0.11$^{*}$ & 0.98$\pm$0.03$^{*}$ & 0.26$\pm$0.05$^{*}$ & 0.24$\pm$0.05$^{*}$ & 0.27$\pm$0.06$^{*}$ & 33.40$\pm$2.14 \\
 K2-106 & 5522$\pm$34 & 4.34$\pm$0.05 & 0.87$\pm$0.05 & 0.10$\pm$0.03 & 0.07$\pm$0.05 & 0.05$\pm$0.03 & 35.25$\pm$1.36 \\
 K2-229 & 5196$\pm$35 & 4.39$\pm$0.07 & 0.91$\pm$0.06 & -0.06$\pm$0.02 & -0.07$\pm$0.03 & -0.06$\pm$0.05 & 33.44$\pm$1.28 \\
 Kepler-107 & 5958$\pm$37 & 4.21$\pm$0.05 & 1.25$\pm$0.04 & 0.42$\pm$0.03 & 0.41$\pm$0.11 & 0.37$\pm$0.04 & 34.79$\pm$2.41 \\
 Kepler-406 & 5625$\pm$28 & 4.26$\pm$0.05 & 0.96$\pm$0.04 & 0.24$\pm$0.02 & 0.23$\pm$0.05 & 0.25$\pm$0.04 & 33.05$\pm$1.31 \\
 K2-291 & 5541$\pm$24 & 4.39$\pm$0.06 & 0.97$\pm$0.04 & 0.09$\pm$0.02 & 0.10$\pm$0.03 & 0.04$\pm$0.04 & 34.10$\pm$1.03 \\
 55 Cnc & 5341$\pm$62 & 4.26$\pm$0.14 & 0.94$\pm$0.09 & 0.32$\pm$0.04 & 0.42$\pm$0.06 & 0.35$\pm$0.05 & 30.22$\pm$1.91 \\
 EPIC 249893012 & 5552$\pm$27 & 4.01$\pm$0.05 & 1.11$\pm$0.03 & 0.15$\pm$0.02 & 0.21$\pm$0.05 & 0.17$\pm$0.03 & 31.33$\pm$1.17 \\
 HD 213885 & 5885$\pm$15 & 4.39$\pm$0.03 & 1.06$\pm$0.02 & 0.00$\pm$0.01 & -0.01$\pm$0.04 & -0.01$\pm$0.03 & 33.95$\pm$0.97 \\
 HD 219134 & 4789$\pm$54 & 4.15$\pm$0.19 & 0.69$\pm$0.14 & -0.03$\pm$0.03 & -0.03$\pm$0.06 & 0.0$\pm$0.06 & 32.45$\pm$1.68 \\
 HD 3167 & 5306$\pm$36 & 4.35$\pm$0.07 & 0.69$\pm$0.06 & 0.04$\pm$0.02 & 0.10$\pm$0.04 & 0.03$\pm$0.04 & 32.15$\pm$1.27 \\
 K2-141 & 4486$\pm$115 & 4.23$\pm$0.35 & 0.62$\pm$0.41 & -0.06$\pm$0.05 & -0.16$\pm$0.11 & -0.03$\pm$0.12 & 34.33$\pm$3.51 \\
 K2-216 & 4505$\pm$172 & 4.21$\pm$0.47 & 0.61$\pm$0.50 & -0.15$\pm$0.09 & -0.23$\pm$0.15 & -0.15$\pm$0.16 & 34.68$\pm$5.23 \\
 K2-265 & 5466$\pm$26 & 4.36$\pm$0.05 & 0.77$\pm$0.05 & 0.09$\pm$0.02 & 0.11$\pm$0.03 & 0.05$\pm$0.04 & 33.73$\pm$1.15 \\
 Kepler-10 & 5669$\pm$16 & 4.33$\pm$0.03 & 0.92$\pm$0.02 & -0.14$\pm$0.01 & -0.01$\pm$0.03 & -0.1$\pm$0.03 & 29.19$\pm$0.82 \\
 Kepler-20 & 5502$\pm$35 & 4.40$\pm$0.06 & 0.83$\pm$0.06 & 0.05$\pm$0.03 & 0.07$\pm$0.04 & 0.06$\pm$0.03 & 32.42$\pm$1.26 \\
 Kepler-78 & 4993$\pm$72 & 4.43$\pm$0.20 & 0.97$\pm$0.16 & -0.14$\pm$0.04 & -0.15$\pm$0.10 & -0.09$\pm$0.07 & 31.80$\pm$2.40 \\
 Kepler-93 & 5620$\pm$17 & 4.41$\pm$0.03 & 0.72$\pm$0.03 & -0.14$\pm$0.01 & -0.14$\pm$0.04 & -0.16$\pm$0.02 & 33.64$\pm$0.96 \\
 TOI-402 & 5081$\pm$43 & 4.21$\pm$0.13 & 0.71$\pm$0.10 & 0.03$\pm$0.03 & 0.09$\pm$0.04 & 0.06$\pm$0.05 & 30.87$\pm$1.38 \\
 WASP-47 & 5559$\pm$52 & 4.32$\pm$0.12 & 1.09$\pm$0.07 & 0.40$\pm$0.04 & 0.46$\pm$0.06 & 0.43$\pm$0.05 & 30.79$\pm$1.86 \\
 TOI-561 & 5314$\pm$20 & 4.37$\pm$0.04 & 0.54$\pm$0.05 & -0.39$\pm$0.02 & -0.18$\pm$0.04 & -0.27$\pm$0.03 & 25.23$\pm$1.06 \\
 Sun & 5777$\pm$10$^{\dag}$ & 4.43$\pm$0.02$^{\dag}$ & 0.95$\pm$0.02$^{\dag}$ & 7.50$\pm$0.04$^{\ddag}$ & 7.60$\pm$0.04$^{\ddag}$ & 7.51$\pm$0.03$^{\ddag}$ & 33.20$\pm$1.70
 \\
\hline
\end{tabular}
}
\\
{\raggedright * The parameters and abundances from \cite{Toledo-Padron-20}. \par}
{\raggedright \dag The solar atmospheric parameters from \cite{Adibekyan-16}. \par}
{\raggedright \ddag The solar absolute abundances from \cite{Asplund-09}. \par}
\end{table}

\begin{table}[ht]
\caption{\label{tab:planets} \textbf{Table S4. Properties of the rocky planets.} The masses, radii, and semi-major axis (in astronomical units, au) of the planets are taken from exoplanet.eu database \cite{Schneider-11}. Columns from 5 to 10 are derived in this work as described in Methods.}
\tiny
{\renewcommand{\arraystretch}{1.7}
\setlength{\tabcolsep}{2.5pt}
\begin{tabular}{lccccccccc}
\hline
             Planet &      $R$ ($R_{\mathrm{\oplus}}$)  &   $M$ ($M_{\mathrm{\oplus}}$) &  Semi-major axis (au) &   $T_{\mathrm{eq}}$ (K) & $\rho / \rho_{Earth-like}$ &   $f_{\mathrm{core}}^{\mathrm{planet}}$ (\%) &  $f_{\mathrm{core\&mantle}}^{\mathrm{planet}}$ (\%) & Fe/(Mg+Si)$_{\mathrm{core}}^{\mathrm{planet}}$ &  Fe/(Mg+Si)$_{\mathrm{core\&mantle}}^{\mathrm{planet}}$\\
\hline
 \multicolumn{9}{c}{super-Mercuries}\\

          K2-38 b  &   1.540$_{-0.140}^{+0.140}$  &   7.31$_{-1.11}^{+1.11}$  & 0.04994 & 1229 & 1.35$\pm$0.42  &   59$_{-23}^{+19}$  &   64$_{-17}^{+17}$ & 1.29$_{-0.78}^{+1.91}$ & 1.96$_{-1.03}^{+2.69}$ \\ 
         K2-106 b  &   1.524$_{-0.157}^{+0.157}$  &   8.36$_{-0.95}^{+0.95}$  & 0.01160 & 2439 &       1.54$\pm$0.51  &   66$_{-21}^{+18}$  &   71$_{-18}^{+16}$ & 1.72$_{-1.00}^{+2.74}$ & 2.63$_{-1.45}^{+4.28}$ \\ 
         K2-229 b  &   1.164$_{-0.066}^{+0.048}$  &   2.59$_{-0.43}^{+0.43}$  & 0.01289 & 1978 &       1.41$\pm$0.33  &   64$_{-17}^{+14}$  &   69$_{-15}^{+13}$ & 1.61$_{-0.81}^{+1.68}$ & 2.39$_{-1.16}^{+2.61}$ \\ 
     Kepler-107 c  &   1.597$_{-0.026}^{+0.026}$  &   9.39$_{-1.77}^{+1.77}$  & 0.06040 & 1422 &       1.45$\pm$0.28  &    69$_{-11}^{+8}$  &    70$_{-10}^{+8}$ & 1.99$_{-0.78}^{+1.06}$ & 2.57$_{-0.91}^{+1.31}$ \\ 
     Kepler-406 b  &   1.435$_{-0.034}^{+0.034}$  &     6.36$_{-1.4}^{+1.4}$  &           0.03600\ddag  & 1541 &       1.50$\pm$0.35  &   72$_{-13}^{+10}$  &    73$_{-11}^{+9}$ & 2.25$_{-1.01}^{+1.66}$ & 2.99$_{-1.24}^{+2.13}$ \\ 

 \multicolumn{9}{c}{super-Earths}\\
         
         55 Cnc e  &   1.947$_{-0.038}^{+0.038}$  &   8.59$_{-0.43}^{+0.43}$  & 0.01544 & 2105 &       0.75$\pm$0.06  &      7$_{-5}^{+6}$  &     14$_{-6}^{+7}$ & 0.07$_{-0.04}^{+0.07}$ & 0.14$_{-0.07}^{+0.09}$ \\ 
 EPIC 249893012 b  &   1.950$_{-0.090}^{+0.090}$  &   8.75$_{-1.09}^{+1.09}$  & 0.04700 & 1567 &       0.76$\pm$0.14  &   16$_{-10}^{+14}$  &   29$_{-12}^{+13}$ & 0.17$_{-0.12}^{+0.21}$ & 0.38$_{-0.20}^{+0.32}$ \\ 
      HD 213885 b  &   1.745$_{-0.052}^{+0.052}$  &   8.83$_{-0.66}^{+0.66}$  & 0.02012 & 2099 &       1.06$\pm$0.12  &    40$_{-10}^{+9}$  &     45$_{-8}^{+8}$ & 0.60$_{-0.21}^{+0.27}$ & 0.84$_{-0.27}^{+0.38}$ \\ 
      HD 219134 b  &   1.502$_{-0.057}^{+0.057}$  &   4.27$_{-0.34}^{+0.34}$  & 0.03700 & 1067 &       0.96$\pm$0.13  &   30$_{-12}^{+12}$  &   39$_{-11}^{+11}$ & 0.39$_{-0.19}^{+0.27}$ & 0.64$_{-0.26}^{+0.41}$ \\ 
      HD 219134 c  &   1.415$_{-0.049}^{+0.049}$  &   3.96$_{-0.34}^{+0.34}$  & 0.06200 & 824 &       1.08$\pm$0.15  &   42$_{-12}^{+11}$  &   48$_{-10}^{+10}$ & 0.64$_{-0.26}^{+0.37}$ & 0.95$_{-0.34}^{+0.51}$ \\ 
        HD 3167 b  &   1.704$_{-0.179}^{+0.146}$  &   5.02$_{-0.38}^{+0.38}$  & 0.01815 & 1870 &       0.74$\pm$0.24  &   31$_{-19}^{+25}$  &   48$_{-19}^{+20}$ & 0.38$_{-0.26}^{+0.60}$ & 0.89$_{-0.51}^{+1.11}$ \\ 
         K2-141 b  &   1.513$_{-0.050}^{+0.050}$  &   5.09$_{-0.41}^{+0.41}$  &           0.00716$^{\dag}$  & 2000 &       1.07$\pm$0.14  &   43$_{-11}^{+11}$  &     48$_{-9}^{+9}$ & 0.67$_{-0.25}^{+0.36}$ & 0.95$_{-0.33}^{+0.48}$ \\ 
         K2-216 b  &   1.805$_{-0.202}^{+0.202}$  &   7.91$_{-1.59}^{+1.59}$  & 0.02800 & 1098 &       0.89$\pm$0.35  &   33$_{-21}^{+27}$  &   49$_{-19}^{+21}$ & 0.45$_{-0.33}^{+0.89}$ & 0.98$_{-0.58}^{+1.48}$ \\ 
         K2-265 b  &   1.710$_{-0.101}^{+0.101}$  &   6.54$_{-0.84}^{+0.84}$  & 0.03376 & 1330 &       0.90$\pm$0.20  &   29$_{-16}^{+17}$  &   41$_{-14}^{+14}$ & 0.36$_{-0.23}^{+0.39}$ & 0.70$_{-0.35}^{+0.63}$ \\ 
         K2-291 b  &   1.589$_{-0.095}^{+0.072}$  &   6.49$_{-1.16}^{+1.16}$  & 0.03261 & 1424 &       1.12$\pm$0.25  &   47$_{-16}^{+14}$  &   53$_{-13}^{+13}$ & 0.79$_{-0.39}^{+0.62}$ & 1.18$_{-0.52}^{+0.87}$ \\ 
         Kepler-10 b  &   1.469$_{-0.030}^{+0.020}$  &   3.33$_{-0.49}^{+0.49}$  & 0.01685 & 2299 &       0.85$\pm$0.14  &   20$_{-11}^{+11}$  &   29$_{-10}^{+10}$ & 0.22$_{-0.13}^{+0.18}$ & 0.38$_{-0.17}^{+0.24}$ \\ 
      Kepler-20 b  &   1.869$_{-0.066}^{+0.034}$  &   9.69$_{-1.41}^{+1.44}$  & 0.04630 & 1196 &       0.92$\pm$0.17  &   28$_{-13}^{+13}$  &   37$_{-11}^{+11}$ & 0.35$_{-0.19}^{+0.27}$ & 0.58$_{-0.25}^{+0.38}$ \\ 
      Kepler-78 b  &    1.20$_{-0.090}^{+0.090}$  &   1.69$_{-0.41}^{+0.41}$  & 0.01000 & 2116 &       0.90$\pm$0.30  &   32$_{-20}^{+24}$  &   47$_{-18}^{+18}$ & 0.41$_{-0.30}^{+0.69}$ & 0.91$_{-0.52}^{+1.10}$ \\ 
      Kepler-93 b  &   1.483$_{-0.025}^{+0.025}$  &    4.0$_{-0.67}^{+0.67}$  & 0.05300 & 1150 &       0.95$\pm$0.17  &   30$_{-13}^{+11}$  &   38$_{-10}^{+10}$ & 0.38$_{-0.19}^{+0.25}$ & 0.61$_{-0.25}^{+0.33}$ \\ 
        TOI-402 b  &   1.699$_{-0.062}^{+0.059}$  &   7.21$_{-0.79}^{+0.79}$  & 0.05245 & 1011 &       0.99$\pm$0.15  &   33$_{-13}^{+12}$  &   42$_{-11}^{+10}$ & 0.45$_{-0.21}^{+0.30}$ & 0.70$_{-0.29}^{+0.42}$ \\ 
        WASP-47 e  &   1.872$_{-0.135}^{+0.135}$  &   9.22$_{-0.99}^{+0.99}$  & 0.01730 & 2127 &       0.89$\pm$0.21  &   31$_{-17}^{+18}$  &   43$_{-15}^{+16}$ & 0.40$_{-0.25}^{+0.47}$ & 0.75$_{-0.39}^{+0.77}$ \\ 
        TOI-561 b  &   1.423$_{-0.066}^{+0.066}$  &   1.59$_{-0.36}^{+0.36}$  & 0.01055 & 2327 &       0.51$\pm$0.14  &    11$_{-8}^{+13}$  &   23$_{-11}^{+13}$ & 0.11$_{-0.08}^{+0.17}$ & 0.28$_{-0.16}^{+0.28}$ \\ 
 
 \multicolumn{9}{c}{Solar System rocky planets}\\

 Mercury  &  0.383$^{*}$  & 0.055 & 0.38700 & 450 & 1.53 &  60$_{-1}^{+1}$  &  64$_{-1}^{+1}$ &  &  \\ 
 Venus  &  0.949$^{*}$  & 0.815 & 0.72300 & 329 & 0.96 &  27$_{-1}^{+1}$  &  30$_{-1}^{+1}$ &  &  \\ 
 Earth  & 1.000 & 1.000 & 1.00000 & 280 & 1 &  30$_{-1}^{+1}$  &  35$_{-1}^{+1}$ &  &  \\ 
 Mars  &  0.532$^{*}$  & 0.107 & 1.52300 & 227 & 1.16 &  20$_{-1}^{+1}$  &  28$_{-1}^{+1}$ &  & \\
 \hline
\end{tabular}
}
\\
{\raggedright \dag The semi-major axis for K2-141~b is from \cite{Barragan-18}. \par}
{\raggedright \ddag The semi-major axis for Kepler-406~b is from \cite{Butler-06}. \par}
{\raggedright * One-bar surface radius of the Solar System planets\cite{Archinal-18}. \par}
\end{table}

\begin{table}[ht]
\caption{\label{tab:planets_2} \textbf{Table S5. Alternative masses and radii of the rocky planets.} These measurements are taken from the NASA Exoplanet Archive\cite{NEA} and are used to investigate the robustness of our results to the heterogeneity in the planet data.}
\scriptsize
{\renewcommand{\arraystretch}{1.9}
\begin{tabular}{l|llc|llc|llc|}
\hline
Planet & $R^{\ast}$ ($R_{\mathrm{\oplus}}$) & $M^{\ast}$ ($M_{\mathrm{\oplus}}$) & Reference$^{\ast}$ & $R^{\dag}$ ($R_{\mathrm{\oplus}}$) & $M^{\dag}$ ($M_{\mathrm{\oplus}}$) & Reference$^{\dag}$ & $R^{\ddag}$ ($R_{\mathrm{\oplus}}$) & $M^{\ddag}$ ($M_{\mathrm{\oplus}}$) & Reference$^{\ddag}$ \\ 
\hline

Kepler-78 b & 1.100$_{-0.100}^{+0.100}$ & 1.97$_{-0.53}^{+0.53}$ & \cite{Stassun-17} & 1.228$_{-0.019}^{+0.018}$ & 1.77$_{-0.25}^{+0.24}$ & \cite{Dai-19} & 1.200$_{-0.090}^{+0.09}$ & 1.87$_{-0.26}^{+0.27}$ & \cite{Grunblatt-15} \\ 
Kepler-10 b & 1.489$_{-0.021}^{+0.023}$ & 3.57$_{-0.53}^{+0.51}$ & \cite{Dai-19} & 1.470$_{-0.030}^{+0.030}$ & 3.72$_{-0.42}^{+0.42}$ & \cite{Weiss-16} &  &  &  \\ 
55 Cnc e & 1.910$_{-0.080}^{+0.080}$ & 8.08$_{-0.31}^{+0.31}$ & \cite{Demory-16} & 1.897$_{-0.046}^{+0.044}$ & 7.74$_{-0.30}^{+0.37}$ & \cite{Dai-19} &  &  &  \\ 
K2-141 b & 1.493$_{-0.035}^{+0.041}$ & 5.16$_{-0.34}^{+0.35}$ & \cite{Dai-19} & 1.540$_{-0.090}^{+0.100}$ & 5.31$_{-0.46}^{+0.46}$ & \cite{Barragan-18} &  &  &  \\ 
WASP-47 e & 1.810$_{-0.027}^{+0.027}$ & 6.83$_{-0.66}^{+0.66}$ & \cite{Vanderburg-17} & 1.773$_{-0.048}^{+0.049}$ & 6.91$_{-0.83}^{+0.81}$ & \cite{Dai-19} &  &  &  \\ 
K2-229 b & 1.197$_{-0.048}^{+0.045}$ & 2.49$_{-0.43}^{+0.42}$ & \cite{Dai-19} &  &  &  &  &  &  \\ 
HD 219134 c & 1.511$_{-0.047}^{+0.047}$ & 4.36$_{-0.22}^{+0.22}$ & \cite{Gillon-17} &  &  &  &  &  &  \\ 
K2-291 b & 1.582$_{-0.042}^{+0.037}$ & 6.40$_{-1.10}^{+1.10}$ & \cite{Dai-19} &  &  &  &  &  &  \\ 
Kepler-93 b & 1.600$_{-0.100}^{+0.100}$ & 4.54$_{-0.84}^{+0.84}$ & \cite{Stassun-17} &  &  &  &  &  &  \\ 
HD 219134 b & 1.606$_{-0.086}^{+0.086}$ & 4.34$_{-0.44}^{+0.44}$ & \cite{Motalebi-15} &  &  &  &  &  &  \\ 
HD 3167 b & 1.626$_{-0.054}^{+0.048}$ & 5.59$_{-0.96}^{+0.98}$ & \cite{Dai-19} &  &  &  &  &  &  \\ 
K2-216 b & 1.750$_{-0.100}^{+0.170}$ & 8.00$_{-1.60}^{+1.60}$ & \cite{Persson-18} &  &  &  &  &  &  \\ 
Kepler-20 b & 1.910$_{-0.210}^{+0.120}$ & 8.70$_{-2.20}^{+2.10}$ & \cite{Gautier-12} &  &  &  &  &  &  \\ 

 \hline
\end{tabular}
}
\\
{\raggedright * First alternative masses and radii of the planets, and the corresponding references.\par}
{\raggedright \dag Second alternative masses and radii of the planets, and the corresponding references. \par}
{\raggedright \ddag Third alternative masses and radii of the planets, and the corresponding references. \par}
\end{table}

\begin{table}[ht]
\caption{\label{tab:statistics} \textbf{Table S6. Results of the ODR and $t$-test for the full sample.}  These correlations are plotted in Figures 1 and 2.}
\begin{center}
{\renewcommand{\arraystretch}{1.3}
\begin{tabular}{lcccc}
\hline
 Data & Slope & Intercept & Residual Variance & P($t$-stat) \\
\hline
$\rho / \rho_{\mathrm{Earth-like}}$ vs $f_{\mathrm{iron}}^{\mathrm{star}}$ & 0.074$\pm$0.012 & -1.4$\pm$0.4 & 0.45 & 7$\times$10$^{-6}$ \\
$f_{\mathrm{core}}^{\mathrm{planet}}$ vs $f_{\mathrm{iron}}^{\mathrm{star}}$ & 6.8$\pm$1.3 & -184$\pm$40 & 0.79 & 2$\times$10$^{-5}$ \\
$f_{\mathrm{core\&mantle}}^{\mathrm{planet}}$ vs $f_{\mathrm{iron}}^{\mathrm{star}}$ & 6.3$\pm$1.2 & -160$\pm$37 & 0.74 & 2$\times$10$^{-5}$ \\
\hline
\end{tabular}
}
\end{center}
\end{table}

\begin{table}[ht]
\caption{\label{tab:statistics_superEarths} \textbf{Table S7. Results of the ODR and $t$-test for just  the Super Earths.} These correlations are plotted in Figures 1 and 2.}
\begin{center}
{\renewcommand{\arraystretch}{1.3}
\begin{tabular}{lcccc}
\hline
 Data & Slope & Intercept & Residual Variance & P($t$-stat) \\
\hline
$\rho / \rho_{\mathrm{Earth-like}}$ vs $f_{\mathrm{iron}}^{\mathrm{star}}$ & 0.061$\pm$0.009 & -1.0$\pm$0.3 & 0.26 & 7$\times$10$^{-6}$ \\
$f_{\mathrm{core}}^{\mathrm{planet}}$ vs $f_{\mathrm{iron}}^{\mathrm{star}}$ & 4.5$\pm$0.8 & -115$\pm$26 & 0.39 & 6$\times$10$^{-5}$ \\
$f_{\mathrm{core\&mantle}}^{\mathrm{planet}}$ vs $f_{\mathrm{iron}}^{\mathrm{star}}$ & 4.3$\pm$0.8 & -100$\pm$27 & 0.42 & 1$\times$10$^{-4}$ \\
\hline
\end{tabular}
}
\end{center}
\end{table}

\begin{table}[ht]
\caption{\label{tab:statistics_tests} \textbf{Table S8. Dependencies of  $\rho / \rho_{\mathrm{Earth-like}}$  and $f_{\mathrm{iron}}^{\mathrm{planet}}$ on the elemental abundances for the full sample.}}
\begin{center}
{\renewcommand{\arraystretch}{1.3}
\begin{tabular}{lcccc}
\hline
 Data & Slope & Intercept & Residual Variance & P($t$-stat) \\
\hline
$\rho / \rho_{\mathrm{Earth-like}}$ vs [Fe/H] & -0.04$\pm$0.20 & 0.88$\pm$0.05 & 1.70 & 0.80 \\
$\rho / \rho_{\mathrm{Earth-like}}$ vs [Mg/H] & -0.24$\pm$0.17 & 0.92$\pm$0.05 & 1.57 & 0.18 \\
$\rho / \rho_{\mathrm{Earth-like}}$ vs [Si/H] & -0.15$\pm$0.20 & 0.90$\pm$0.05 & 1.67 & 0.47 \\
\\
$f_{\mathrm{core}}^{\mathrm{planet}}$ vs [Fe/H] & 8.2$\pm$21.8 & 30.8$\pm$5.0 & 3.33 & 0.71 \\
$f_{\mathrm{core}}^{\mathrm{planet}}$ vs [Mg/H] & -28.8$\pm$20.4 & 35.5$\pm$5.3 & 3.16 & 0.17 \\
$f_{\mathrm{core}}^{\mathrm{planet}}$ vs [Si/H] & -6.8$\pm$22.8 & 32.3$\pm$5.3 & 3.34 & 0.77 \\
\\
$f_{\mathrm{core\&mantle}}^{\mathrm{planet}}$ vs [Fe/H] & 14.5$\pm$19.7 & 41.9$\pm$4.1 & 2.74 & 0.47 \\
$f_{\mathrm{core\&mantle}}^{\mathrm{planet}}$ vs [Mg/H] & -13.5$\pm$18.9 & 44.4$\pm$4.4 & 2.76 & 0.48 \\
$f_{\mathrm{core\&mantle}}^{\mathrm{planet}}$ vs [Si/H] & 4.6$\pm$21.2 & 42.6$\pm$4.3 & 2.8 & 0.83 \\
\hline
\end{tabular}
}
\end{center}
\end{table}

\newpage

\clearpage
\newpage

\textbf{Data S1.} A python code to determine the iron mass fraction using the stoichiometric model described in \cite{Santos-15, Santos-17}. The input parameters of the code are the relative-to-the-sun abundances of Mg ([Mg/H]), Si ([Si/H]), and Fe ([Fe/H]). The code works with python2.7 and python3.

\end{document}